\documentclass[a4paper,fleqn,usenatbib,useAMS]{mnras}

\usepackage[T1]{fontenc}
\usepackage{ae,aecompl}

\usepackage{graphicx}	% Including figure files
\usepackage{stackengine}
\usepackage{epstopdf}
\usepackage{savesym}
\usepackage{amsmath}
\savesymbol{iint}
\savesymbol{iiint}
\usepackage{txfonts}
\restoresymbol{TXF}{iint}
\restoresymbol{TXF}{iiint}
\usepackage{fix-cm}
\usepackage{subfig}
\usepackage[]{units}
\usepackage{textcomp}

\allowdisplaybreaks


\title[The role of zonal flows in disc gravito-turbulence]{The role of zonal flows in disc gravito-turbulence}

\author[R. Vanon]{
R. Vanon\thanks{E-mail: \href{mailto:r.vanon@sheffield.ac.uk}{r.vanon@sheffield.ac.uk}}
\\
DAMTP, CMS, University of Cambridge, Wilberforce Road, Cambridge, CB3 0WA, UK\\
SoMaS, University of Sheffield, Hicks Building, Hounsfield Rd, Sheffield, S3 7RH, UK
}

\date{Accepted XXX. Received YYY; in original form ZZZ}

\pubyear{2017}

\begin{document}
\label{firstpage}
\pagerange{\pageref{firstpage}--\pageref{lastpage}}
\maketitle

% Abstract of the paper
\begin{abstract}
  The work presented here focuses on the role of zonal flows in the self-sustenance of gravito-turbulence in accretion discs. The numerical analysis is conducted using a bespoke pseudo-spectral code in fully compressible, non-linear conditions. The disc in question, which is modelled using the shearing sheet approximation, is assumed to be self-gravitating, viscous, and thermally diffusive; a constant cooling timescale is also considered. Zonal flows are found to emerge at the onset of gravito-turbulence and they remain closely linked to the turbulent state. A cycle of zonal flow formation and destruction is established, mediated by a slow mode instability (which allows zonal flows to grow) and a non-axisymmetric instability (which disrupts the zonal flow), which is found to repeat numerous times. It is in fact the disruptive action of the non-axisymmetric instability to form new leading and trailing shearing waves, allowing energy to be extracted from the background flow and ensuring the self-sustenance of the gravito-turbulent regime.
\end{abstract}

% Select between one and six entries from the list of approved keywords.
% Don't make up new ones.

%%%%%%%%%%%%%%%%%%%%%%%%%%%%%%%%%%%%%%%%%%%%%%%%%%

%%%%%%%%%%%%%%%%% BODY OF PAPER %%%%%%%%%%%%%%%%%%

\section{Introduction}
Disc self-gravity (SG) can play a crucial role in the dynamics of accretion discs in AGN discs or early stage protoplanetary (PP) discs; it can allow an outward transfer of angular momentum or result in disc fragmentation, a process which has been linked with the formation of stellar/sub-stellar or planetary bodies. The (inverse) strength of a disc's SG is usually quantified by means of the Toomre parameter \citep{Toomre1964}

\begin{equation}
  Q \equiv \frac{c_s \kappa}{\pi G \Sigma},
\end{equation}
where $c_s$ represents the sound speed, $\kappa$ the epicyclic frequency, $G$ is the gravitational constant and $\Sigma$ the surface density.
The temperature of the disc strongly affects the outcome of gravitational instability (GI), with cold discs being more susceptible to its onset. Using the simple $\beta$-cooling prescription, \citet{Gammie2001} showed that disc fragmentation would be triggered if the cooling timescale $\tau_c$ of the disc obeyed

\begin{equation}
  \beta \equiv \tau_c \Omega \lesssim 3,
\end{equation}
although more recent numerical works have found somewhat different threshold values \citep{Riceetal2003, Riceetal2005}. Furthermore, \citet{Paardekooper2012} found the fragmentation process to be of a stochastic nature, with its probability simply decreasing with longer cooling times. 

A less efficient cooling time than $\beta \simeq 3$, on the other hand, results in the development of a self-sustaining gravito-turbulent state, where the Toomre parameter is maintained at roughly $Q\sim1$ by the opposing actions of shock dissipation and cooling. The self-sustenance of this turbulent state, a state which can be thought of as a sub-critical instability, however requires a continuous energy extraction from the background flow, the origin of which is still unclear. This work tries to address the question of gravito-turbulence self-sustenance by exploring the possibility of zonal flows being involved in the process. 

Zonal flows are coherent structures of axisymmetric nature exhibiting alternating bands, with an axisymmetric slow mode instability believed to be involved in their formation in discs \citep{VanonOgilvie2017}. They represent equilibrium solutions to the equations of disc flow dynamics in the presence of a geostrophic balance between the pressure gradient and the Coriolis force. 
Zonal flows occur frequently in astrophysical fluids; the closest example is Earth's atmosphere, where the resulting sharp temperature gradients may trigger cyclones and precipitation. They are also observed in the atmospheres of other Solar System planets, the most notable example being Jupiter's striped pattern. 

Zonal flows have also been observed in simulations of MHD turbulent accretion discs \citep{Johansenetal2009, Simonetal2012, KunzLesur2013, BaiStone2014} performed using the shearing box approximation; these works found the emergence of zonal flows to be independent of both initial conditions and box size. The incompressible inviscid hydrodynamical calculation by \citet{Lithwick2007, Lithwick2009} -- carried out using the shearing sheet model in non-SG conditions -- showed that zonal flows can be broken up into vortices by the action of the Kelvin-Helmholtz (KH) instability. These vortices, observed by several other simulations \citep[e.g.][]{UmurhanRegev2004, JohnsonGammie2005}, appear to be long-lived despite the modest Reynolds numbers that can be applied in numerical simulations of accretion discs. More refined work on the stability of zonal flows by \citet{VanonOgilvie2016} -- carried out in compressible and SG conditions -- found that, as well as being affected by the KH instability as identified by \citet{Lithwick2007}, zonal flows can also be gravitationally unstable for $Q \lesssim 2$. 

%{\color{red}While zonal flows could play a major role in the dynamics of protoplanetary discs by slowing down or halting protoplanets' potentially destrucrive migration due to gas drag during the `metre-sized barrier' phase,} 
The aim of this paper is to investigate whether zonal flows play a central role in the self-sustenance of gravito-turbulence, and how this self-sustaining process is maintained. This analysis is carried out thanks to a 2D pseudo-spectral code specially written for this work; the code makes use of the shearing sheet model, with the modelled flow being fully compressible, viscous and self-gravitating. A simple $\beta$-cooling prescription is also employed, as well as a horizontal thermal diffusion. Section~\ref{sec:casper} introduces the equations solved by the pseudo-spectral method employed, together with its specifications. The main results are presented in Section~\ref{sec:results}, with particular emphasis given to the mechanism of self-sustenance of the turbulent state; the implications of these findings and the concluding remarks are presented in Section~\ref{sec:discussion}.

\section{The \texttt{CASPER} code} \label{sec:casper}
For the purpose of this analysis, a pseudo-spectral code based on the shearing sheet model was developed. The code, which was named \texttt{CASPER}, takes into account the self-gravity of the disc and solves fully compressible, viscous non-linear equations for the evolution of the flow. \texttt{CASPER} makes use of a third order Runge-Kutta iteration method, representing the best compromise between accuracy and performance. 

Several reasons were at the root of the decision to employ a pseudo-spectral code to tackle this problem. One of these is the obvious selling point of spectral methods' accuracy, which allows for a much faster (i.e. exponential) error convergence than other methods. Furthermore, spectral methods' affinity to systems with periodic boundary conditions (which are applied in both $x$ and $y$ in this case) and ease to deal with disc self-gravity represented another two strong advantages of this choice. The method's ability to resolve and analyse each individual wavelength independently was also a useful tool for this work, allowing a close comparison to previous zonal flow stability analyses \citep{VanonOgilvie2016, VanonOgilvie2017}.
Of course spectral methods do come with their own weaknesses; in particular, and most appropriately for this problem, their difficulty to appropriately resolve shocks. The problem was however circumvented with the use of viscosities, as explained in more detail in Section~\ref{sec:specs}.

\subsection{The shearing sheet model}
The \texttt{CASPER} code is based upon the local shearing sheet model first used by \citet{GoldreichLynden-Bell1965} in the context of galactic discs. The model employs a Cartesian frame of reference with periodic boundary conditions for both spatial coordinates, and it is centred around the fiducial radius $R_0$. The sheet, which has dimensions $L_x$ and $L_y$ obeying $L_x$, $L_y \ll R_0$, corotates with the disc. In this corotating frame of reference, a viscous, compressible flow is described by the continuity and Navier-Stokes equations:

\begin{equation} \label{eq:continuity-sigma}
	\frac{\partial \Sigma}{\partial t} + \nabla \cdot \left(\Sigma \boldsymbol{\varv}\right) = 0,
\end{equation}
\begin{equation}
	\frac{\partial \boldsymbol{\varv}}{\partial t} + \boldsymbol{\varv} \cdot \nabla \boldsymbol{\varv} + 2\boldsymbol{\Omega} \times \boldsymbol{\varv} = - \nabla \Phi - \nabla \Phi_{d,m} - \frac{1}{\Sigma}\nabla P - \frac{1}{\Sigma}\nabla \cdot \boldsymbol{T},
\end{equation}
where $\boldsymbol{\varv}$ is the flow velocity vector, $\boldsymbol{\Omega} = \Omega \boldsymbol{e}_z$ the angular velocity of the disc ($\boldsymbol{e}_z$ being the unit vector parallel to the $z$-axis), $\Phi = -q \Omega^2 x^2$ the effective potential, $q=-\mathrm{d}\ln \Omega/\mathrm{d}\ln r$ the dimensionless shear rate (its value being $q=3/2$ for Keplerian discs), $\Phi_{d,m}$ the disc potential being evaluated at the disc midplane and P the 2-dimensional pressure. Also, 

\begin{equation}
	\boldsymbol{T} = 2\mu_s \boldsymbol{S} + \mu_b \left(\nabla \cdot \boldsymbol{\varv}\right) \boldsymbol{I}
\end{equation}
represents the stress tensor for the shear ($\mu_s$) and bulk ($\mu_b$) dynamic viscosities ($\mu_i = \Sigma \nu_i$, with $\nu_i$ being the corresponding kinematic viscosity), $\boldsymbol{I}$ is the unit tensor and $\boldsymbol{S}$ represents the traceless shear tensor, which is given by

\begin{equation}
	\boldsymbol{S} = \frac{1}{2} \left[ \nabla \boldsymbol{\varv} + \left(\nabla \boldsymbol{\varv}\right)^T \right] - \frac{1}{3} \left(\nabla \cdot \boldsymbol{\varv}\right) \boldsymbol{I}.
\end{equation}

The continuity equation (Equation~\ref{eq:continuity-sigma}) is transformed, by means of the introduction of the quantity $h=\ln \Sigma + \mathrm{const}$, to

\begin{equation}
	\frac{\partial h}{\partial t} + \boldsymbol{\varv}\cdot \nabla h + \nabla \cdot \boldsymbol{\varv} = 0.
\end{equation}

The self-gravity of the disc is regulated by the $\nabla \Phi_\mathrm{d,m}$ term; this can be evaluated using the Poisson equation

\begin{equation}
	\nabla^2 \Phi_\mathrm{d} = 4\pi G\Sigma \delta (z),
\end{equation}
with $\delta(z)$ representing the Dirac delta function and $z$ being the height from the disc midplane. The solution to the equation is most easily expressed in Fourier space, with the full form of the disc potential being

\begin{equation}
	\tilde{\Phi}_\mathrm{d} = - \frac{2\pi G \tilde{\Sigma}}{\sqrt{k_x^2 + k_y^2}} \mathrm{e}^{- \lvert \boldsymbol{k}\rvert \lvert z\rvert},
\end{equation}
where $\tilde{\Sigma}$ represents the Fourier transform of the respective quantity, and with $k_x$ and $k_y$ being the radial and azimuthal wavenumbers. While this expression is a function of the height from the disc midplane $z$, the midplane form of the disc potential can be readily found by setting $z=0$, obtaining

\begin{equation}
	\tilde{\Phi}_\mathrm{d,m} = - \frac{2\pi G\tilde{\Sigma}}{\sqrt{k_x^2 + k_y^2}}.
\end{equation}

Other important quantities in the analysis include the potential vorticity $\zeta$ and the specific entropy $s$, which are given by:

\begin{equation}
	\zeta = \frac{2\Omega + \left(\nabla \times \boldsymbol{\varv}\right)_z}{\Sigma},
\end{equation}
\begin{equation}
	s = \frac{1}{\gamma} \ln P - \ln \Sigma,
\end{equation}
where $\gamma$ represents the adiabatic index (which is taken as\footnote{Although $\gamma=2$ does not necessarily represent the most physically realistic scenario, it offers a direct comparison with much of the literature, which have adopted this value after \citet{Gammie2001}. } $\gamma=2$ in this analysis) and the pressure $P$ is given by

\begin{equation}
	P = (\gamma-1) \Sigma e;
\end{equation}
here $e$ represents the specific internal energy, whose temporal evolution is dictated by

\begin{equation}
	\frac{\partial e}{\partial t} + \boldsymbol{\varv}\cdot \nabla e = - \frac{P}{\Sigma} \nabla \cdot \boldsymbol{\varv} + 2\nu_s \boldsymbol{S}^2 + \nu_b \left( \nabla \cdot \boldsymbol{\varv}\right)^2 + \frac{1}{\Sigma} \nabla \cdot \left( \nu_t \Sigma \nabla e\right) - \frac{e}{\tau_c}.
\end{equation}
Three types of diffusive effects feature in the equation: bulk ($\nu_b$) and shear ($\nu_s$) viscosities, and (horizontal) thermal diffusion ($\nu_t$); a constant $\beta$-cooling time $\tau_c$ is also considered.

During the simulation runs, the flow evolves away from its background state. It is therefore useful to express each quantity as a sum of its background state and its departure away from it, e.g. $\Sigma = \Sigma_0 + \Sigma^\prime$ (with $\Sigma_0$ representing the background state value and $\Sigma^\prime$ the departure), or $\boldsymbol{\varv} = \boldsymbol{\varv}_0 + \boldsymbol{\varv}^\prime$ (with $\boldsymbol{\varv}_0 = (0,-q\Omega x,0)^T$ and $\boldsymbol{\varv}^\prime = (u^\prime,\varv^\prime,0)^T$). The departure from the background state can then be described by the following set of equations:

\begin{equation}
	\mathrm{D}h^\prime = - \left(\partial_x u^\prime + \partial_y \varv^\prime\right) \equiv - \Delta,
\end{equation}
\begin{multline}
	\mathrm{D}u^\prime - 2\Omega \varv^\prime = - \partial_x \Phi_\mathrm{d,m}^\prime - (\gamma-1) \left(\partial_x e^\prime + e\, \partial_x h^\prime \right) \\+ \nu_s \nabla^2 u^\prime + \left(\nu_b + \frac{1}{3}\nu_s\right) \partial_x \Delta + T_{xx}\partial_x h^\prime + T_{xy} \partial_y h^\prime ,
\end{multline}
\begin{multline}
	\mathrm{D}\varv^\prime + (2-q)\Omega u^\prime = - \partial_y \Phi_\mathrm{d,m}^\prime - (\gamma-1)\left(\partial_y e^\prime + e\, \partial_y h^\prime\right) \\+ \nu_s \nabla^2 \varv^\prime + \left(\nu_b + \frac{1}{3}\nu_s\right) \partial_y \Delta + T_{yx} \partial_x h^\prime + T_{yy} \partial_y h^\prime ,
\end{multline}
\begin{equation} \label{eq:e-full}
	\mathrm{D}e^\prime = - (\gamma-1)e \Delta + \nu_s U + \left(\nu_b - \frac{2}{3}\nu_s\right) \Delta^2 + \nu_t \nabla^2 e^\prime - \frac{e}{\tau_c} .
\end{equation}
Here $\mathrm{D} = \partial_t + u^\prime \partial_x + \varv^\prime \partial_y$ is the Lagrangian derivative, and the above equations have been simplified by the quantities $T_{xx}$, $T_{xy}$, $T_{yx}$, $T_{yy}$ and $U$, which are given by

\begin{equation}
	T_{xx} = 2\nu_s \partial_x u^\prime + \left( \nu_b - \frac{2}{3}\nu_s \right) \Delta,
\end{equation}
\begin{equation}
	T_{yy} = 2\nu_s \partial_y \varv^\prime + \left( \nu_b - \frac{2}{3}\nu_s \right) \Delta,
\end{equation}
\begin{equation}
	T_{xy}=T_{yx} = \nu_s \left( -q\Omega + \partial_x \varv^\prime + \partial_y u^\prime\right),
\end{equation}
\begin{equation}
	U = 2 \left(\partial_x u^\prime\right)^2 + 2\left(\partial_y \varv^\prime\right)^2 + \left(-q\Omega + \partial_x \varv^\prime + \partial_y u^\prime\right)^2.
\end{equation}
It is possible to notice through Equation~\ref{eq:e-full} that the background state of the flow does not represent a steady state solution of the equations. For this reason, as discussed in more detail in Section~\ref{sec:turb-visc} also considering viscous and thermal effects, the initial state is not in thermal equilibrium.

\subsubsection{Stresses}
Another quantity of interest in analysing the flow dynamics is the stress tensor. In particular, its Reynolds (or hydrodynamical) and gravitational components, which can be used to estimate the amount of angular momentum transport $\alpha$. 
We consider spatially averaged gravitational and Reynolds stresses, $\left\langle G_{xy}\right\rangle$ and $\left\langle H_{xy}\right\rangle$ respectively, which are given by

\begin{align}
	\left\langle G_{xy}\right\rangle = & \frac{1}{L_x L_y} \int \int G_{xy} \, \mathrm{d}x \, \mathrm{d}y \nonumber \\ = &- \frac{1}{4\pi G} \sum_k \frac{k_x k_y}{\lvert \boldsymbol{k}\rvert} \left\lvert \tilde{\Phi}_{d,m} (\boldsymbol{k})\right\rvert^2,
\end{align}

\begin{align}
	\left\langle H_{xy} \right\rangle = & \frac{1}{L_x L_y} \int \int H_{xy} \, \mathrm{d}x \, \mathrm{d}y \nonumber \\
  = & \frac{1}{L_x L_y} \int \int \Sigma u \varv \, \mathrm{d}x \, \mathrm{d}y,
\end{align}
where $G_{xy}$ and $H_{xy}$ are the respective local stresses.

Once the stresses have been obtained, the Reynolds stress having been calculated in real space to avoid a Fourier convolution, they can be used to calculate the value of $\alpha$ according to

\begin{equation} \label{eq:stress-alpha}
	\alpha = \frac{\left\langle G_{xy} + H_{xy}\right\rangle}{qP}.
\end{equation}

\subsection{Specifications} \label{sec:specs}
\subsubsection{Diffusive processes}
As mentioned previously, the analysis conducted with \texttt{CASPER} employs three types of diffusive processes: bulk and shear viscosities, and (horizontal) thermal diffusion. All three kinematic diffusion coefficients are taken to be independent of radius, temperature or surface density for reasons of simplicity. This also ensures the flow to be viscously stable, as \citep{LightmanEardley1974}

\begin{equation}
	\frac{\partial \left(\nu \Sigma\right)}{\partial \Sigma} > 0.
\end{equation}

The kinematic diffusion coefficients are therefore initialised as constants and are expected to retain their initial values for the remainder of the simulation. However, if the spatial resolution is not sufficiently high to allow strong shocks to be appropriately resolved using that viscous configuration, the code can increase the viscosity coefficients to avoid numerical errors. This is done by continuously identifying the largest $x$- and $y$-velocities in the flow 

\begin{align}
	U_\mathrm{max} = &\, \lvert u\rvert_\mathrm{max} + c_s,\nonumber \\
	V_\mathrm{max} = &\, \lvert \varv\rvert_\mathrm{max} + c_s,
\end{align}
and checking whether the initial viscosity is larger than the viscosity needed to resolve flows moving at $U_\mathrm{max}$ or $V_\mathrm{max}$ at the given spatial resolution.

Regardless of the existence of this safety measure employed to avoid numerical artefacts such as the Gibbs phenomenon, it is important to stress that steps have been taken to make sure the viscosity coefficients remain constant, with the rare deviations not exceeding $5\%$ of the initial value.

\subsubsection{Time stepping}
The time step of each simulation $\Delta t$ was likewise continuously adapted to the evolving flow to ensure stability according to

\begin{equation}
	\Delta t = \min \left( \tau_\mathrm{visc}, \min\left( \tau_{\mathrm{adv,}x}, \tau_{\mathrm{adv,}y}\right) \right),
\end{equation}
where $\tau_\mathrm{visc}$, $\tau_{\mathrm{adv,}x}$ and $\tau_{\mathrm{adv,}y}$ are the viscous and radial and azimuthal advection timescales, respectively, given by

\begin{align}
	& \tau_\mathrm{visc} \simeq C_\nu \frac{1}{\nu k_\mathrm{max}^2}, \nonumber \\
	& \tau_{\mathrm{adv,}x} = C_\mathrm{CFL} \frac{\Delta x}{U_\mathrm{max}}, \\
	& \tau_{\mathrm{adv,}y} = C_\mathrm{CFL} \frac{\Delta y}{V_\mathrm{max}} \nonumber.
\end{align}
Here $k_\mathrm{max}$ is the largest wavenumber resolved, and $C_\nu$ and $C_\mathrm{CFL}$ are safety factors, the latter being controlled by the Courant-Friedrichs-Lewy (CFL) condition.

\subsubsection{Anti-aliasing}
Another of \texttt{CASPER}'s features is the presence of an anti-aliasing filter, which can be easily turned on or off. This is particularly important in the periodic boundary conditions setting used, as radial wavenumbers exceeding the maximum absolute values set by the run parameters are remapped back on the other side of the $k_x$ range using the classical 
\begin{equation}
k_x(t) = k_x(0) + k_y q\Omega t.
\end{equation} 
This causes trailing waves exceeding the largest resolved wavenumber to be remapped as leading waves and viceversa. For this purpose, \texttt{CASPER} employs a truncation (or 2/3-rule) anti-aliasing method, where shearing waves having wavenumbers exceeding 2/3 the largest resolved wavenumber are discarded. This results in a continuous, but minimal, energy loss.

Although useful, such a method is only able to remove aliasing errors arising from quadratic non-linearities. This means that -- while the method would be able to completely eliminate aliasing errors in an incompressible, non-gravitational flow -- in the gravitational, compressible case errors resulting from cubic or higher order non-linearities remain.

\subsection{Shock resolving test}
Several tests were carried out to ensure the code worked as expected on simplified problems, before tackling the one at hand. One such test was to verify the ability of the code to handle and resolve shocks, given spectral methods' known weakness in dealing with flow discontinuities. 

\begin{figure}
  \includegraphics[width=\columnwidth]{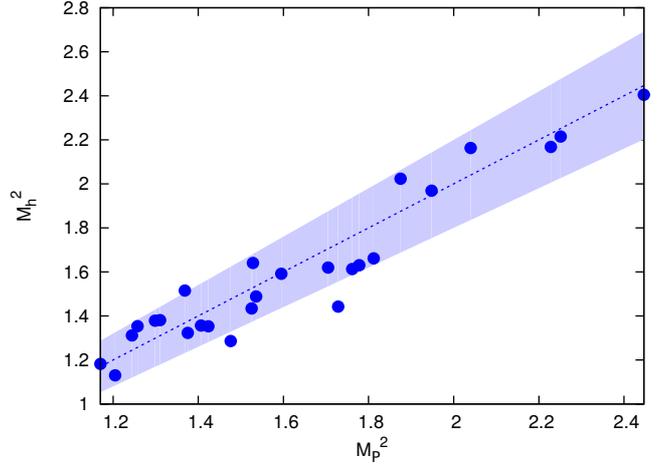}
  \caption{Comparison between the Mach numbers obtained from $h$ and $P$ to test the efficiency of the \texttt{CASPER} code to resolve shocks. The dashed line represents the ideal case where $\mathcal{M}_h = \mathcal{M}_P$, while the shaded area encloses values falling within $10\%$ of this ideal case. The code is seen to handle shocks adequately, with its performance not deteriorating for stronger shocks.}
  \label{fig:shock-test}
\end{figure}

The test was carried out by computing the Mach numbers for specific shocks using both $h$ and the pressure $P$ from the Rankine-Hugoniot conditions. The flow considered was fully self-gravitating, $Q\sim 1$, and with random initial conditions (as further explained in Section~\ref{sec:ic}). In an ideal case, the two Mach number values $\mathcal{M}_h$ and $\mathcal{M}_P$ should of course be equal. Figure~\ref{fig:shock-test} illustrates the result of this basic test; most of the data points lie close to the dashed line, which illustrates the ideal $\mathcal{M}_h=\mathcal{M}_P$. In fact, all but three points lie within the shaded region, which represents values within $10\%$ the ideal case. Furthermore, the accuracy of the code's shock resolving seem unaffected by the strength of the shock, which confirms the ability of the code to deal with the problem at hand.

\subsection{Initial conditions} \label{sec:ic}
Similarly to other works, the runs presented here are initialised with random velocity initial conditions (ICs). The spectrum of the applied velocity perturbations, which obeys a uniform distribution in the range $[-0.5,0.5]$ and is scaled by a factor $10^{-3} c_s$, is bound by a minimum and maximum wavelength according to

\begin{align}
	k_\mathrm{min}=&\,\frac{2\pi}{L},\nonumber \\
	k_\mathrm{max}=&\,32 k_\mathrm{min}.
\end{align} 

Density and internal energy, on the other hand, are kept uniform with their background values being

\begin{align}
	\label{eq:h0} h_0 = &\,\ln \Sigma_0 = 0, \\
	\label{eq:e0} e_0 = &\,\frac{c_s^2}{\gamma(\gamma-1)}.
\end{align}
These ICs aim to mimic early residual disc turbulence following the collapse of its parent cloud \citep{GodonLivio2000}. Unlike other works employing random velocity perturbation ICs \citep{JohnsonGammie2005, Shenetal2006}, no incompressibility condition was applied in this instance.

The runs examined in this work are carried out on a $1024\times 1024$ grid with a square box of dimension $L=8\pi (\pi G\Sigma_0)/\Omega^2$. The box parameters are such that the value of the intrinsic shear viscosity is $\alpha_s\approx 0.004$; as mentioned previously, the code will not increase this value by more than $\sim5\%$. The adiabatic index, as mentioned previously, is set to $\gamma=2$ and a range of the initial Toomre parameter ($1\leq Q_0 \leq 2$) and of the cooling timescale ($7 \leq \tau_c \Omega \leq 15$) are used.

\section{Results} \label{sec:results}
\begin{figure*}
	\centering
	\includegraphics[width=.8\textwidth]{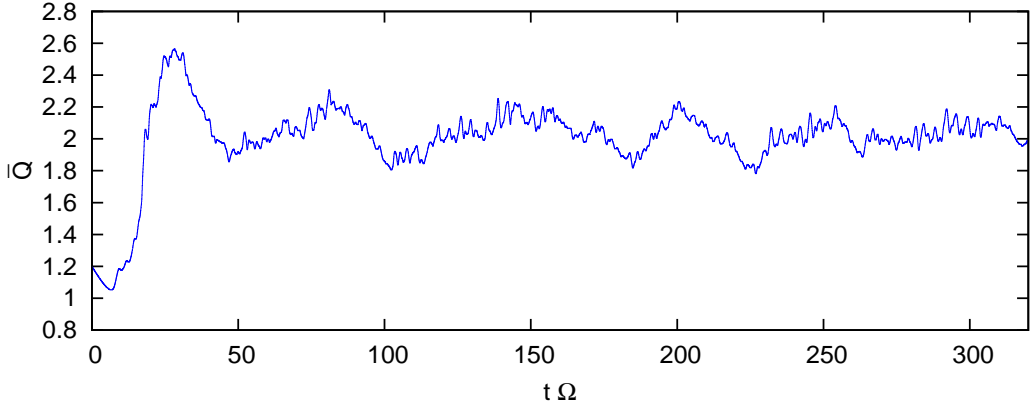}
  \caption{Evolution of the average Toomre parameter $\overline{Q}$ as a function of time with $\tau_c\Omega=12$, showing the system settling into a self-regulated state of gravito-turbulence following an initial period of cooling.}
  \label{fig:Qbar}
\end{figure*}

Having applied the initial conditions mentioned above, the flow was allowed to evolve freely for tens of orbits. The system quickly settled into a self-sustaining state with an average Toomre parameter $\overline{Q} \equiv \sqrt{\gamma (\gamma-1) \bar{e}} \approx 2$ as shown in Figure~\ref{fig:Qbar} (the expression being obtained thanks to the use of gravitational units, set such that $c_s=\pi G\Sigma_0=1$, and of Equation~\ref{eq:e0}), with the quantity $\bar{e}=\tilde{e}\left(k_x=0,k_y=0\right)$ representing the mean internal energy. Runs with different box properties showed no significant difference in the average turbulent $Q$. Owing to the system not being in thermal equilibrium with the applied ICs, the disc is observed to cool at first ($\tau_c\Omega=12$ for this run), until the heat generated by the GI reverses the trend at $t\Omega \approx 7$. As $\overline{Q}$ increases, the amount of viscous heating produced by smoothing down shocks decreases until the system settles into a state of self-regulating gravito-turbulence with $\overline{Q}\approx 2$ starting from $t\Omega \approx 50$. This gravito-turbulent state, which features recurring weaker heating events on a characteristic timescale of $\sim50 \Omega^{-1}$ ($\sim 8$ orbits), persists until the end of the run.

The main question addressed in this work is what mechanism allows the gravito-turbulent state to be self-sustaining. While axisymmetric shearing waves dominate the dynamics for $Q<1$, in this case these are not present as the disc is not sufficiently cool. The shearing sheet model used here only allows non-axisymmetric shearing waves to be transiently amplified. The system is linearly stable to non-axisymmetric perturbations, as in the linear regime viscous effects quickly quench such transient growths. It is however possible for the system to continuously regenerate transiently growing shearing waves by means of a coupling with a non-linear feedback. This coupling would allow to continuously extract energy from the background flow and feed it into the non-axisymmetric GI, causing a sustained state of gravito-turbulence to survive.

\begin{figure}
  \includegraphics[width=\columnwidth]{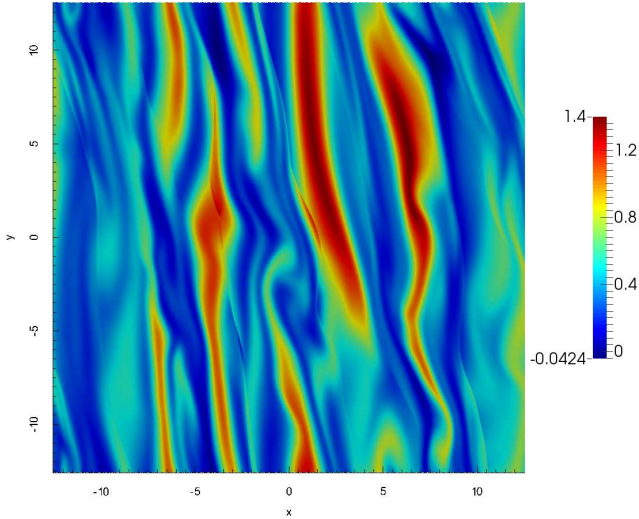}
  \caption{Snapshot of the flow in the entropy showing the presence of the nearly axisymmetric structure.}
  \label{fig:flow_zf}
\end{figure}

\begin{figure*}
  \centering
  \hspace*{-.24cm}\includegraphics[width=145mm]{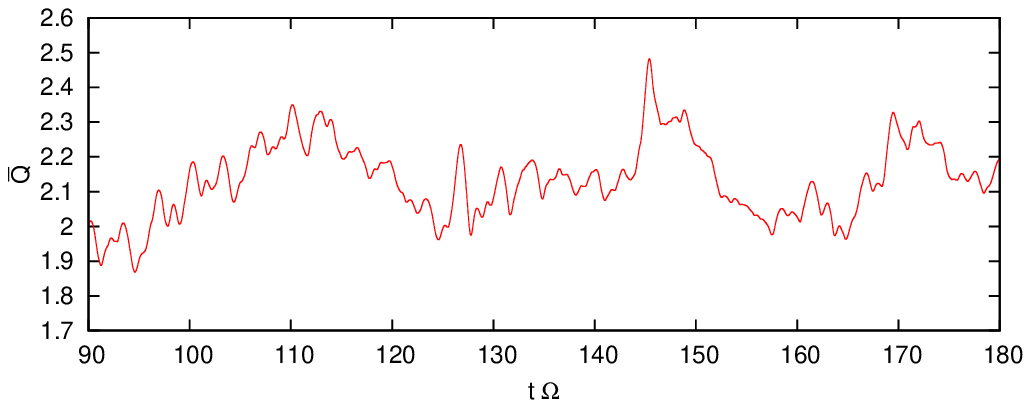}\\
  \hspace*{1.2cm} \includegraphics[width=165mm]{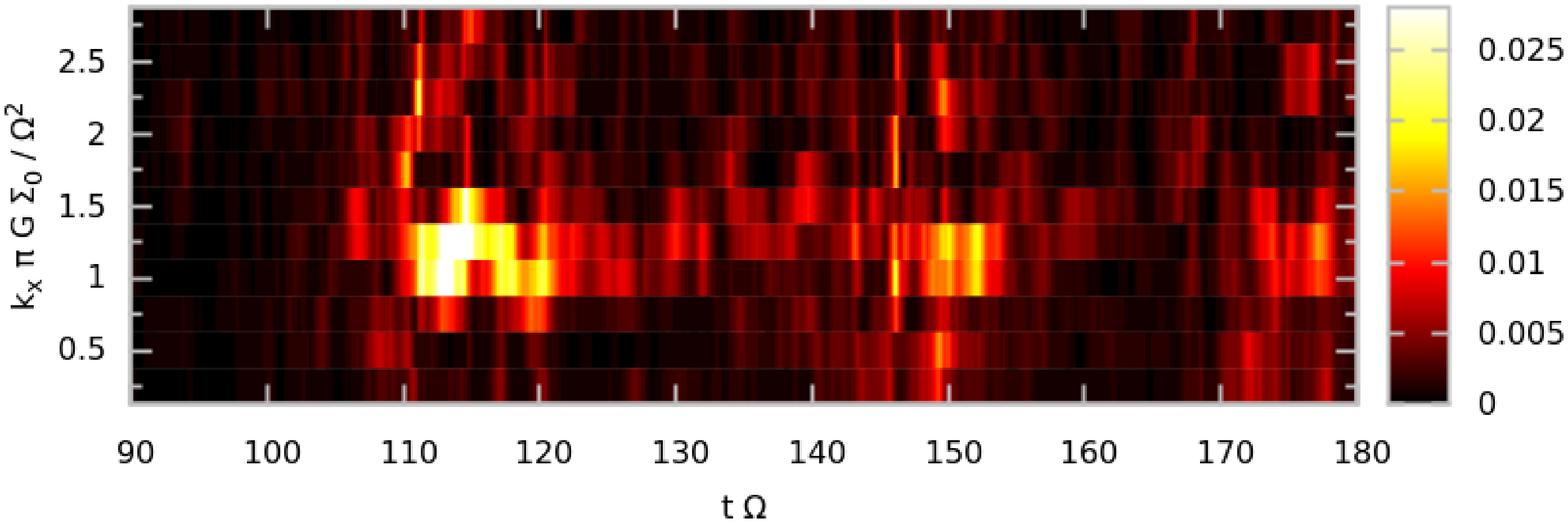}\\
  \hspace*{1.1cm} \includegraphics[width=165mm]{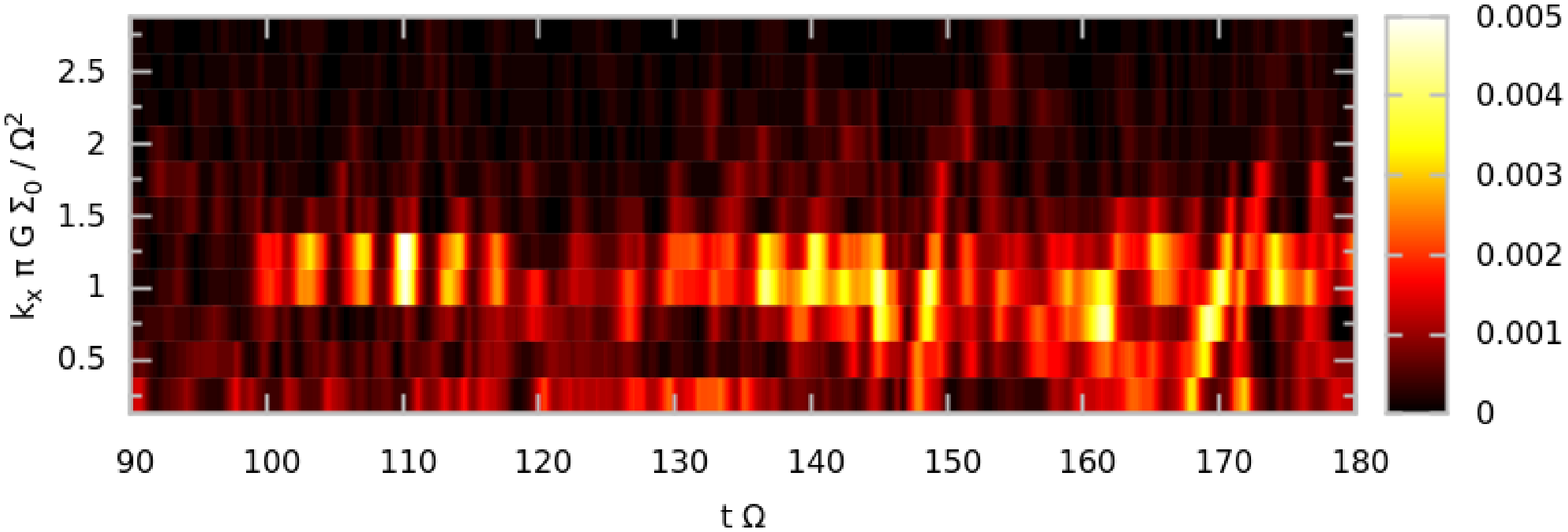}
  \caption{Temporal evolution of $\overline{Q}$ (top) and the axisymmetric power spectrum maps for potential vorticity (middle) and entropy (bottom) in the interval $90 \leq t\Omega \leq 180$ for $\tau_c\Omega=10$. The power maps show the wavenumbers $k_x \pi G\Sigma_0/\Omega^2=1.00$ and $1.25$ dominating over the other components, becoming especially prominent during heating events.}
  \label{fig:2zf}
\end{figure*}

Analysing the behaviour of the flow during the gravito-turbulent state can provide the first hints regarding how such a state is maintained. Figure~\ref{fig:flow_zf} shows the spatial behaviour of the flow in the entropy and features the presence of a nearly axisymmetric structure (henceforth called a zonal flow), which is also present in the potential vorticity. The zonal flow, which in this case is found to have a wavenumber $k_x \pi G\Sigma_0/\Omega^2=1.25$, is found to persist while the system is in its gravito-turbulent state. In fact, the structure is observed to be disrupted and reformed again on a timescale comparable to that of the heating events seen in Figure~\ref{fig:Qbar}. This correlation between the heating events and the zonal flow evolution is illustrated in Figure~\ref{fig:2zf}, showing the time evolution of $\overline{Q}$ (top) as well as the axisymmetric power spectrum maps for PV (middle) and entropy (bottom). Figure~\ref{fig:2zf} shows that there are in fact two dominating axisymmetric $k_x$ values, and that their values evolve with the flow, although they remain close to $k_x \pi G\Sigma_0/\Omega^2 \sim 1$. Runs with different box sizes and lower resolutions, not presented here, confirmed the dominant wavenumbers to possess $k_x \pi G\Sigma_0/\Omega^2 \sim1$, indicating the preference towards these zonal flow wavelengths is dictated by the intrinsic flow behaviour, and not by the box properties.  

\subsection{Turbulent viscosity} \label{sec:turb-visc}
The ability of the flow to self-sustain at a roughly constant value of $Q$, and therefore at a roughly constant temperature (by means of the relationship $c_s \propto T^{1/2}$), indicates that the system has reached a state of thermal equilibrium in its gravito-turbulent regime. However, the intrinsic viscosities and cooling timescale used in the initial conditions do not allow thermal balance. In fact, according to the thermal equilibrium condition

\begin{equation} \label{eq:thermal-eq}
	\alpha_s \tau_c = \frac{1}{q^2 \Omega (\gamma-1)},
\end{equation}
(which is derived by considering only shear viscosity heating and cooling) the initial shear viscosity $\alpha_s \sim 0.004$ would have been thermally balanced by a cooling timescale of only $\tau_c \Omega \sim 100$. Instead, the cooling timescales used are of the order $\tau_c \Omega \sim 10$, which makes the system thermally unbalanced to begin with, which explains the initial cooling period in Figure~\ref{fig:Qbar}.

It is therefore clear that an additional source of viscosity, turbulent in nature, is allowing the system to achieve thermal equilibrium for $\tau_c\Omega \sim 10$. This is confirmed by calculating the effective shear viscosity $\alpha_\mathrm{eff} = \alpha_\mathrm{turb} + \alpha_\mathrm{init}$, where $\alpha_\mathrm{init}$ is the initial shear viscosity and $\alpha_\mathrm{turb}$ the turbulent component given by Equation~\ref{eq:stress-alpha}. The results are found to match very well with the thermal equilibrium condition (Equation~\ref{eq:thermal-eq}).

In the same way that the turbulent motions create an enhanced shear viscosity, the horizontal thermal diffusion would also be boosted. Estimating the turbulent component for $\nu_t$ is however much harder than for $\nu_s$ as \texttt{CASPER} was not designed to compute a detailed heat transport. Instead, the turbulent thermal diffusion is estimated by means of the turbulent Prandtl number
 
\begin{equation}
	\mathrm{Pr}_\mathrm{turb} = \frac{\nu_{s,\mathrm{turb}}}{\nu_{t,\mathrm{turb}}} \sim \frac{H_{xy} + G_{xy}}{H_{xy}}.
\end{equation}
This estimation arises from two assumptions: that fluid motions transport heat and angular momentum in similar ways, and that gravitational interactions carry angular momentum much more efficiently than they transport heat. The value of the turbulent Prandtl number is found to fluctuate around $\mathrm{Pr}\sim 2$.

\subsection{Sustenance through slow mode instability} \label{sec:axi}
\begin{figure*}
  \centering
  \hspace*{-.3mm}\includegraphics[width=140.5mm]{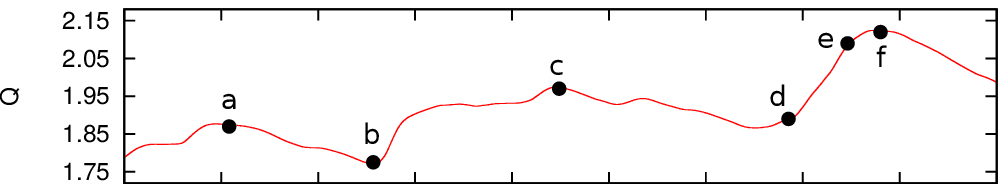}\\
  \includegraphics[width=140mm]{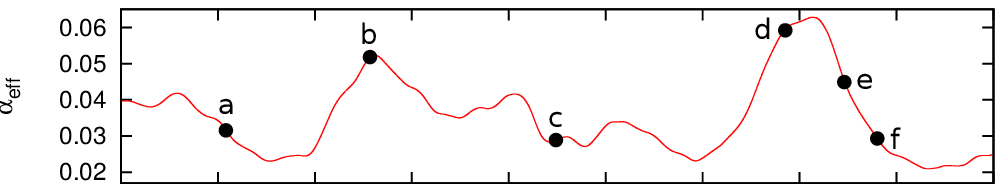}\\
  \hspace*{2.35mm}\includegraphics[width=143.9mm]{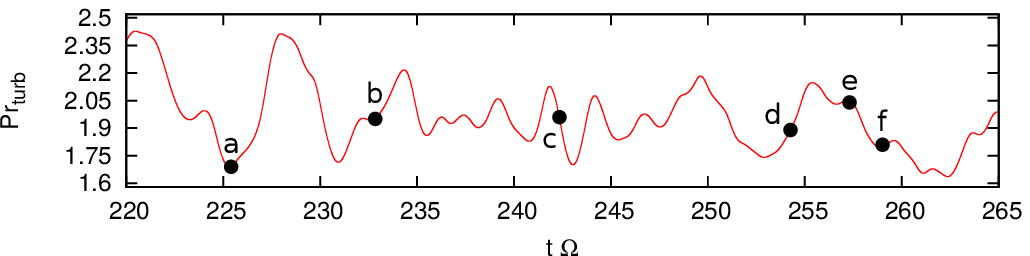}\\
  \vspace*{.5cm}
  \includegraphics[width=48mm]{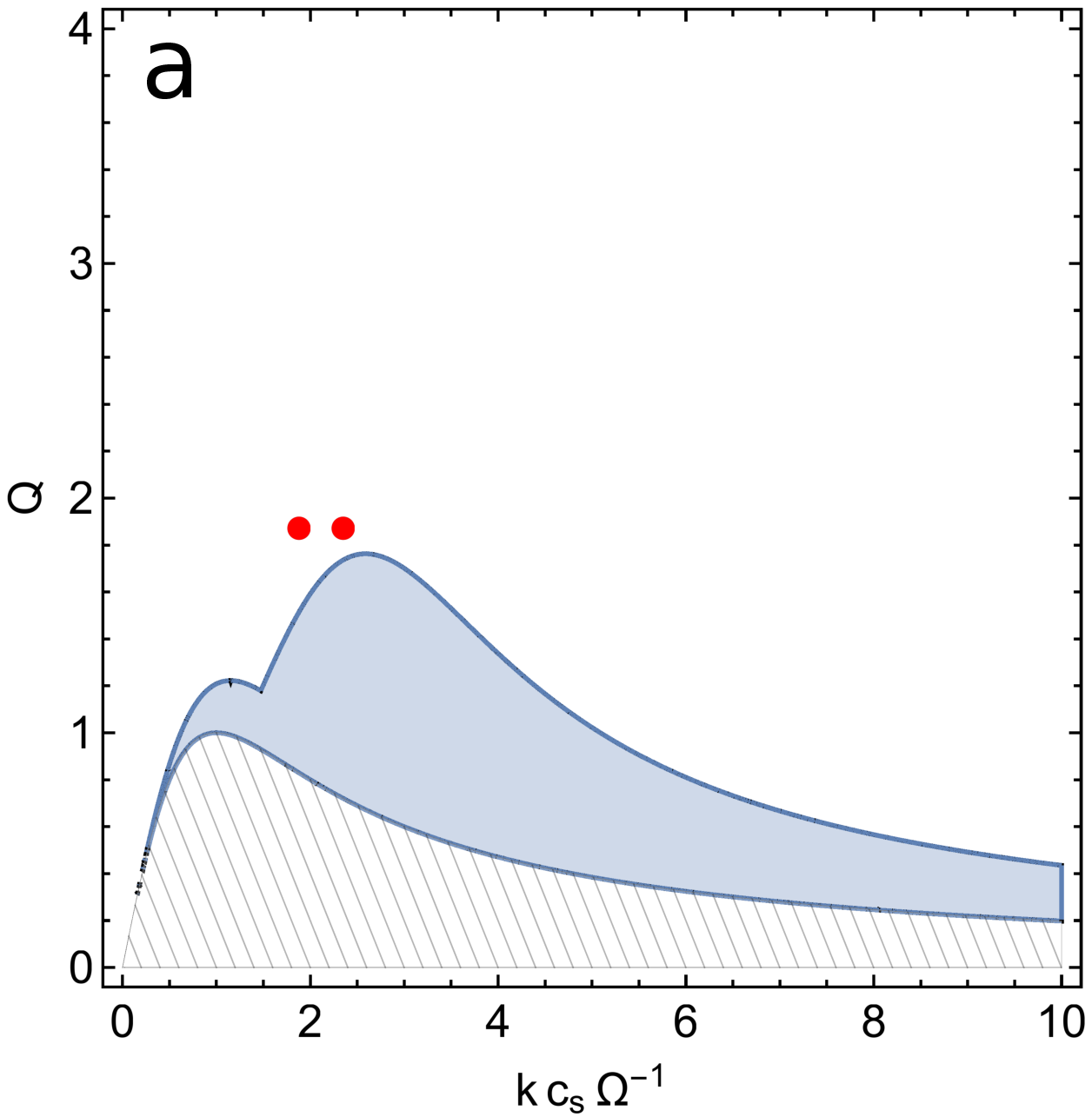}
  \includegraphics[width=48mm]{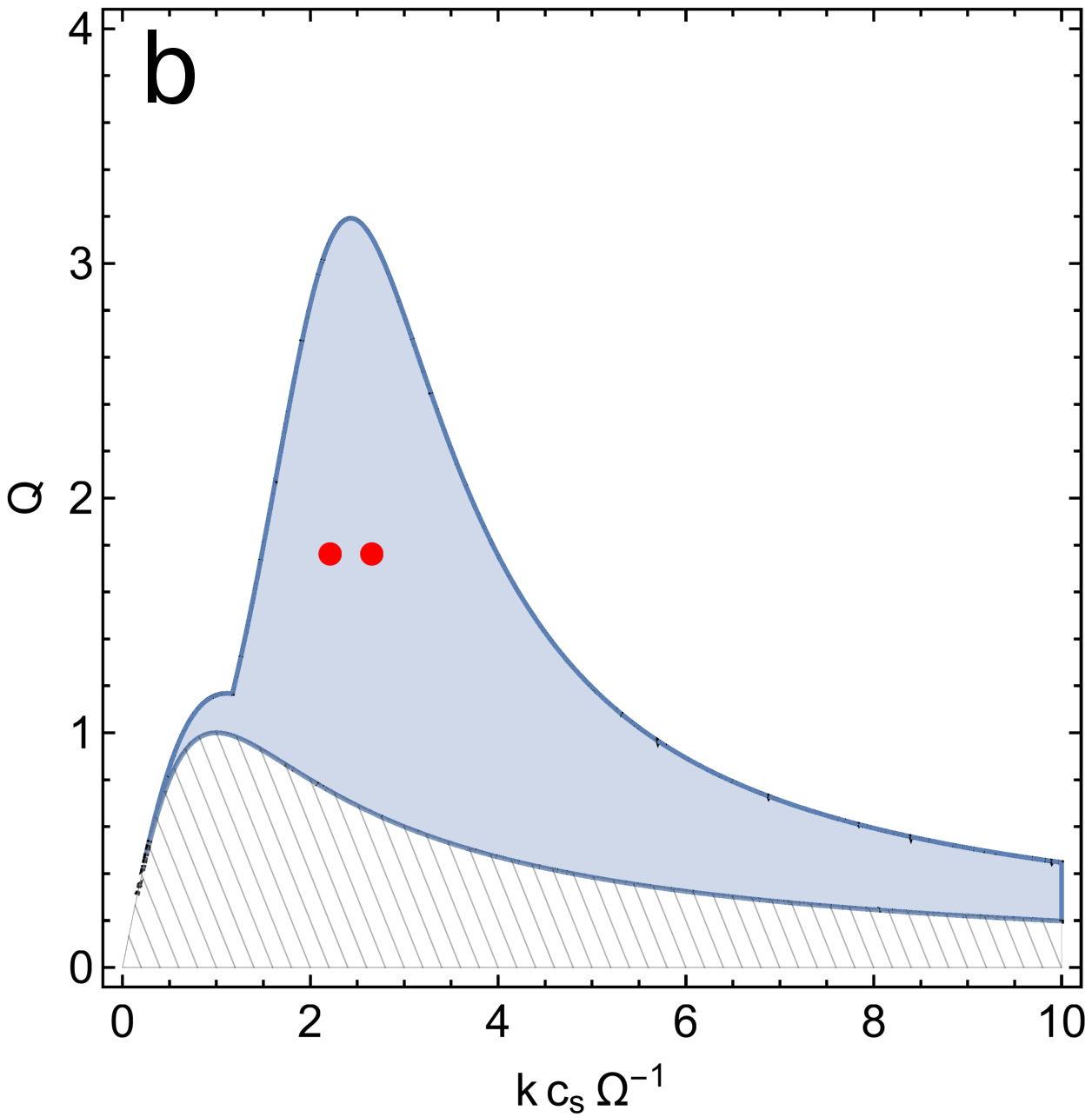}
  \includegraphics[width=48mm]{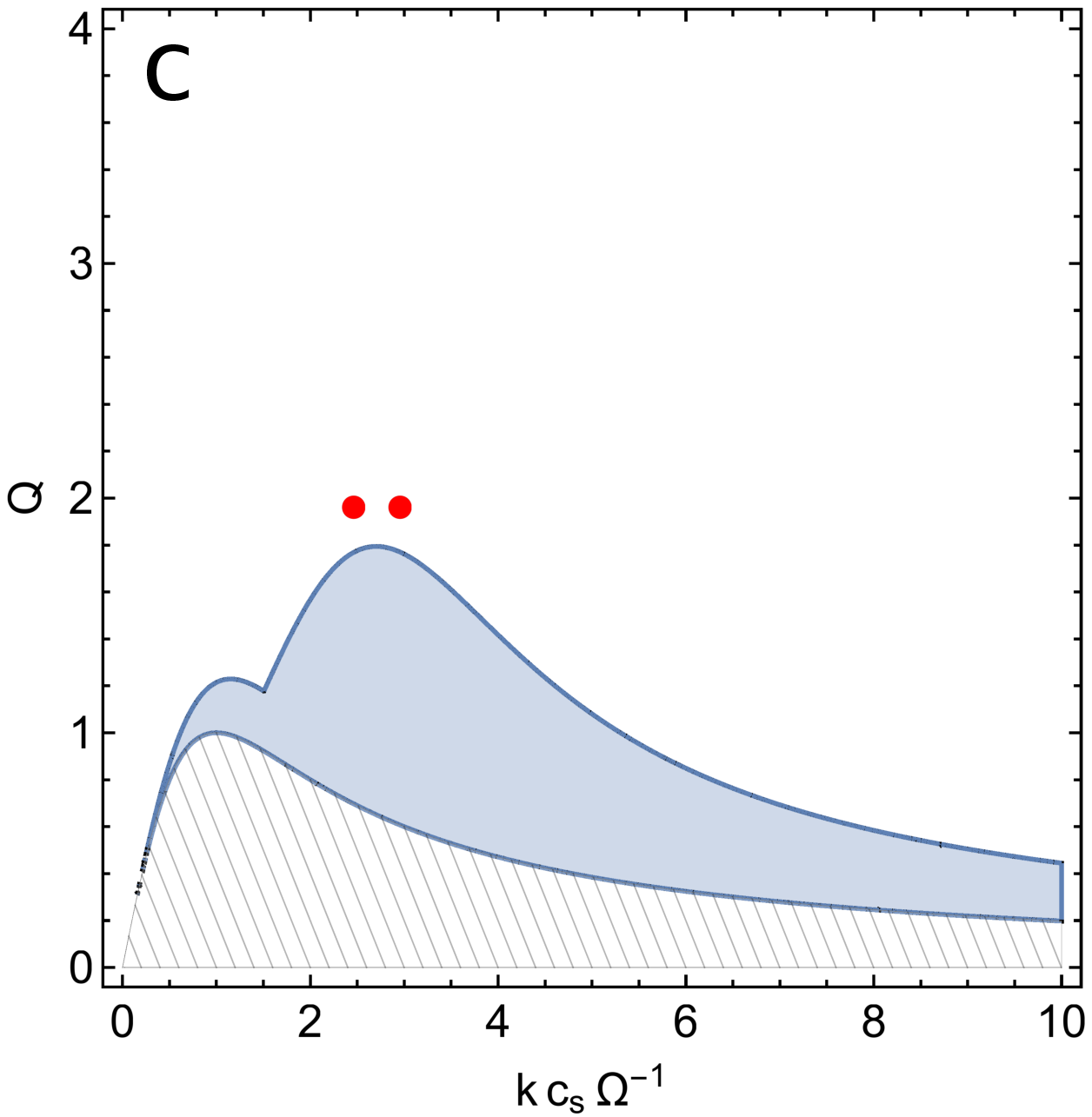}\\
  \includegraphics[width=48mm]{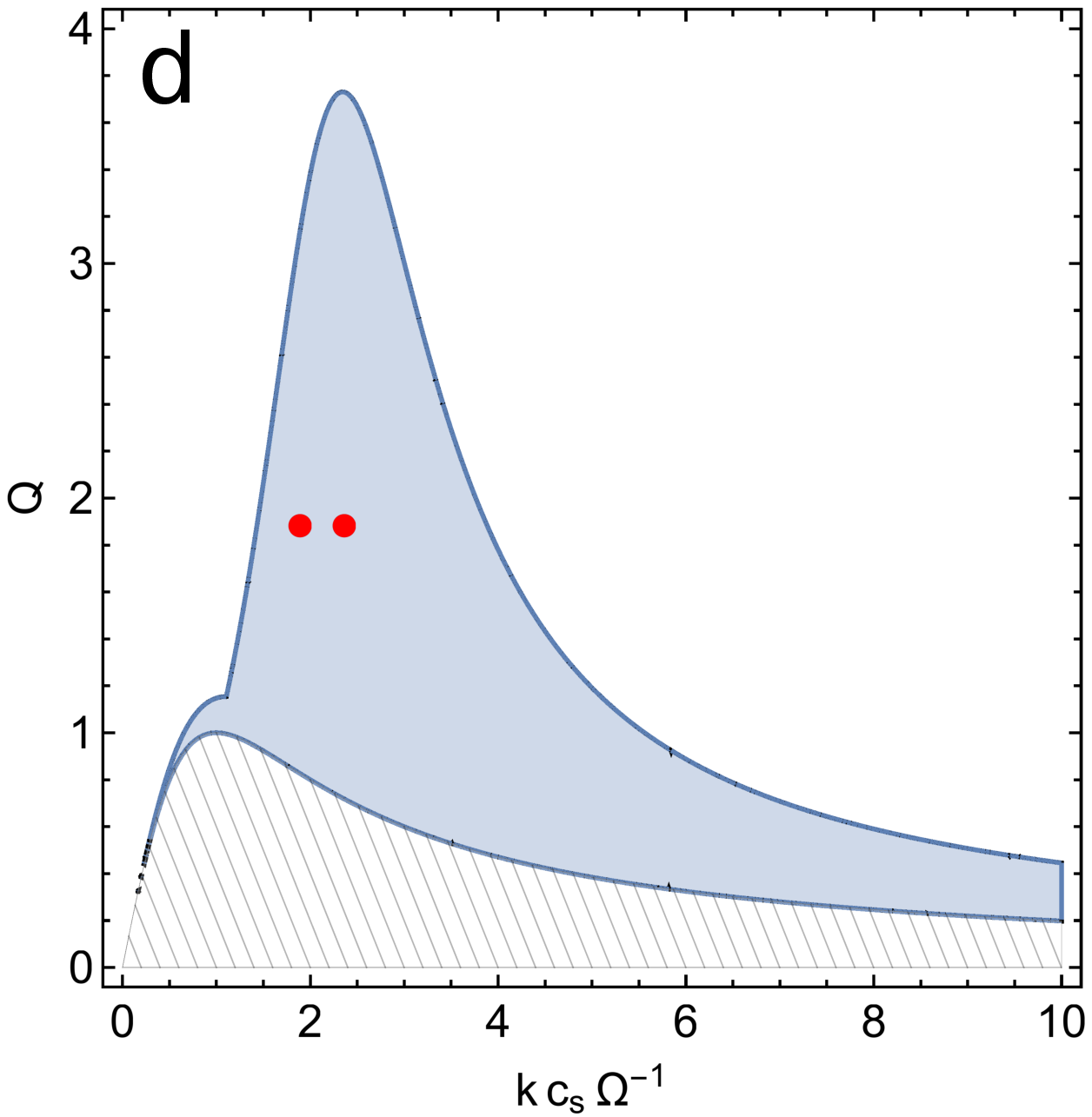}
  \includegraphics[width=48mm]{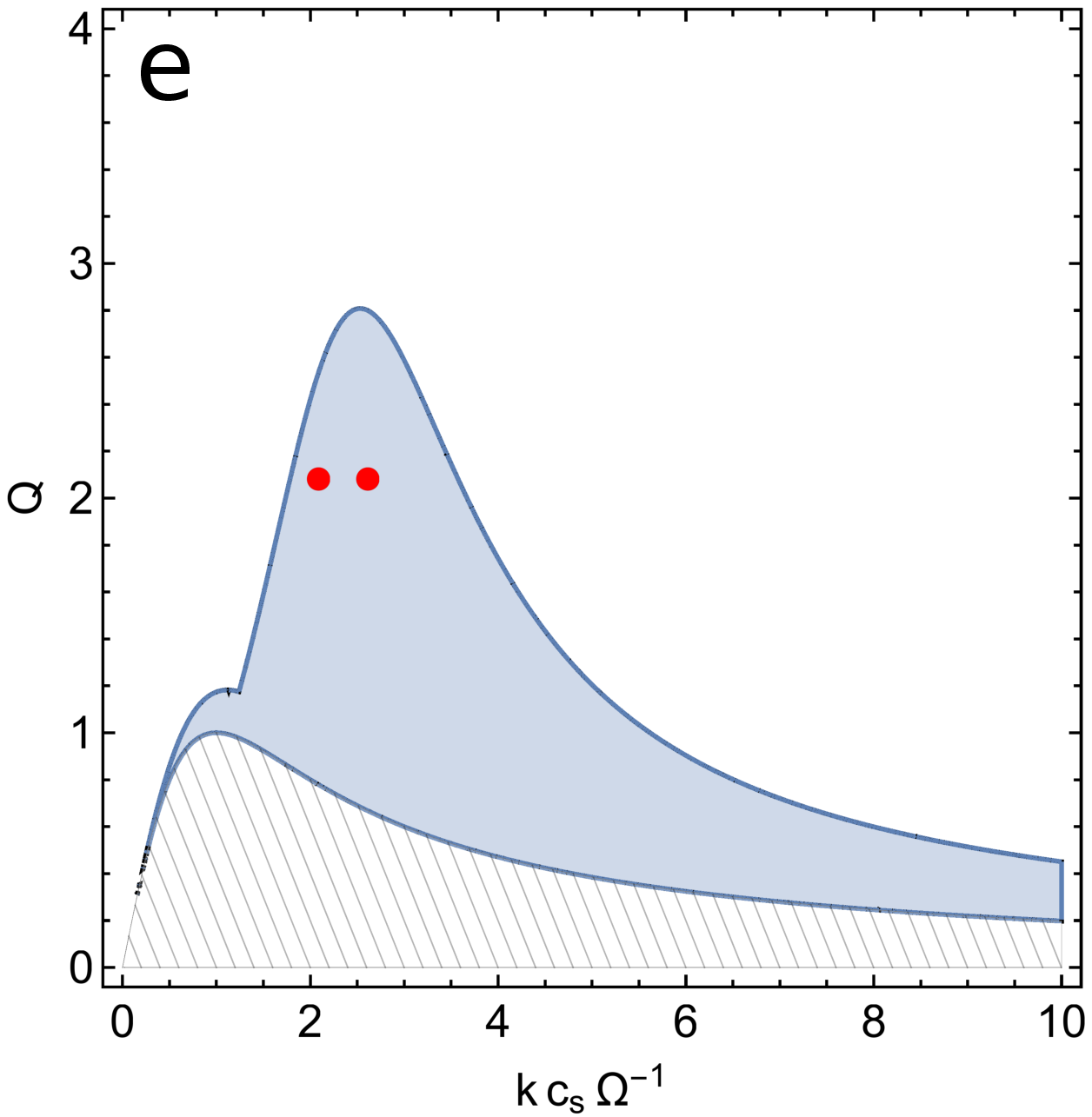}
  \includegraphics[width=48mm]{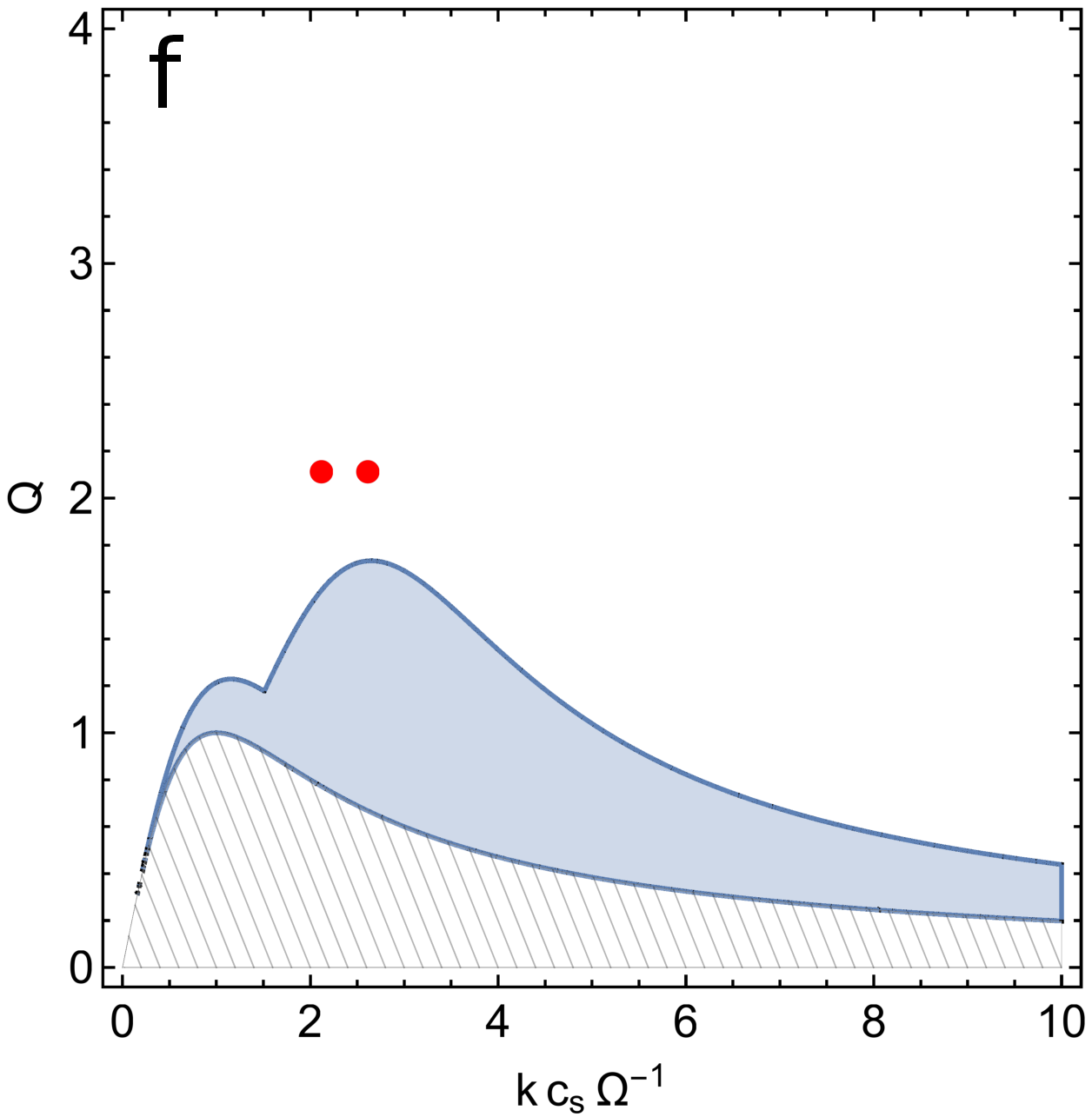}
  \caption{Temporal evolution of $Q$, $\alpha_\mathrm{eff}$ and $\mathrm{Pr}_\mathrm{turb}$ (top) during a heating event including six points marked \textit{a--f} in a run with a cooling timescale $\tau_c \Omega =13$, and the corresponding slow mode instability regions at these snapshots (bottom). The red data points represent the dominating zonal flow wavenumbers in the same run. The shaded area illustrates the region where $\omega_0^2<0$ and the flow is therefore axisymmetrically unstable.}
  \label{fig:axi-casper-grid}
\end{figure*}

As observed in the power spectrum maps of Figure~\ref{fig:2zf} the zonal flows appear to be periodically regenerated through the run, rathen than decaying gradually. This hints at the possibility of an instability acting to grow these axisymmetric structures during heating events.

One plausible candidate is the axisymmetric slow mode instability discussed in \citet{VanonOgilvie2017}, which was shown to generate axisymmetric structures for intermediate wavelengths if the disc was sufficiently cool. The critical temperature was however found to depend on disc parameters such as cooling timescale, effective viscosity, adiabatic index and Prandtl number.
The growth rate $\lambda_\mathrm{sm}$ of such an instability was found to be given by

\begin{equation}
  \lambda_\mathrm{sm} = \frac{c_1 + c_4}{2} \pm \sqrt{\frac{\left(c_1-c_4\right)^2}{4}+c_2c_3},
\end{equation}
where the coefficients $c_1$, $c_2$, $c_3$ and $c_4$ originated from the linearised equations for the zonal flow amplitude's temporal evolution once they were rewritten according to the system

\begin{align} \label{eq:slow-system}
  \partial_t A_s = & \, c_1 A_s + c_2 A_\zeta, \nonumber \\
  \partial_t A_\zeta = & \, c_3 A_s + c_4 A_\zeta,
\end{align}
where $A_s$ and $A_\zeta$ represent the dimensionless zonal flow amplitudes in the specific entropy and in the potential vorticity, respectively, and are given by

\begin{align}
  A_s = & \, \frac{1}{\gamma} \left(A_e + A_h\right), \nonumber \\
  A_\zeta = & \, \frac{k A_v}{(2-q)\Omega} - A_h.
\end{align}
The analysis by \citet{VanonOgilvie2017} found the coefficients appearing in Equation~\ref{eq:slow-system} to be given by:

\begin{align}
	c_1 = & \, \frac{\gamma_t \left(c_s^2 k^2 (\gamma-1) - \gamma \omega_0^2\right) + \gamma_s q \kappa^2 \gamma(\gamma-1)}{\gamma \omega_0^2}, \nonumber \\
	c_2 = & \, \frac{\kappa^2(\gamma-1) \left[\gamma_t c_s^2/\gamma + \gamma_s q/k^2\left(\kappa^2-\omega_0^2\right)\right]}{c_s^2 \omega_0^2} , \nonumber \\
	c_3 = & \, -\frac{4\gamma_s c_s^2 k^2 (q-1)\Omega^2}{\kappa^2 \omega_0^2}, \nonumber \\
	c_4 = & \, -\frac{\gamma_s \left(\omega_0^2 + 4(q-1)\Omega^2\right)}{\omega_0^2}.
\end{align}
Here $\gamma_s = \nu_s k^2$ and $\gamma_t = \nu_t k^2 + 1/\tau_c$ are dissipative coefficients and $\omega_0^2 = \kappa^2 - 2\pi G\Sigma_0 k + c_s^2 k^2$.

To investigate whether the slow mode instability was at the root of the zonal flow growth, the resulting instability region was monitored during a heating event.
Figure~\ref{fig:axi-casper-grid} shows the temporal evolution of the critical varying quantities -- meaning $Q$, $\alpha_\mathrm{eff}$ and $\mathrm{Pr}_\mathrm{turb}$ -- as well as the resulting slow mode instability region in the $kc_s/\Omega$--$Q$ plane at six time points (\textit{a}--\textit{f}) for a run with $\tau_c\Omega =13$. The instability regions also feature red points indicating the two dominating zonal flow wavenumbers (converted into acoustic units) at that particular snapshot. The time sequence shows how the evolution of $\alpha_\mathrm{eff}$ and $\mathrm{Pr}_\mathrm{turb}$ heavily affects the size of the instability region. This is particularly sensitive to $\alpha_\mathrm{eff}$, as shown by peaks in the viscosity (coincident with troughs in $Q$) driving deeply unstable conditions in the disc (snapshots \textit{b}, \textit{d}). The mechanism of the instability is particularly clear in frames \textit{d}--\textit{f}. As the system cools down, the viscosity receives a boost to dissipate strong shocks, driving strongly unstable conditions stretching up to $Q\sim4$ (snapshot \textit{d}). As the disc warms up due to this viscous heating, the value of $\alpha_\mathrm{eff}$ gradually drops, continually shrinking the instability region (\textit{e}), eventually allowing the system to regain stability (\textit{f}).

The slow mode instability playing a part in the self-sustenance of the gravito-turbulent state also explains the presence of two dominating axisymmetric wavenumber modes; in fact, Figure~\ref{fig:axi-casper-grid} shows that these modes' wavenumbers are usually centred around the peak of the instability region, where the growth rate (when the system is unstable) is maximised. While ideally the zonal flow would have the exact wavelength corresponding to the fastest growing mode, the wavenumber quantisation due to the consideration of a box of finite size means two modes (one on either side of the fastest growing mode) are activated instead.

It is however important to remember that some approximations have been made in arriving to this somewhat surprising result. These include the assumption that the turbulent viscosity behaves similarly to the laminar disc viscosity and can therefore be modelled in the same way \citep{BalbusPapaloizou1999}, the rough estimation of the turbulent Prandtl number and the value of $\alpha_\mathrm{eff}$ not being fixed as it was in the slow mode instability analysis by \citet{VanonOgilvie2017}. The former point is the most important but it is believed that the use of a local shearing sheet system, coupled with the gravito-turbulent nature of the disc, should minimise -- if not remove altogether -- the presence of global wave transport, therefore allowing the turbulent disc to be described by the $\alpha$ formalism \citep{BalbusPapaloizou1999}.

Lastly, while it is possible for the intrinsic viscosities to trigger the slow mode instability, their small magnitude ($\alpha \sim 0.004$) means this is unlikely to happen. A confirmation of this is given in Figure~\ref{fig:axi-casper-grid}, where the system fails to trigger the slow mode instability for turbulent shear viscosities of $\alpha_\mathrm{eff} \sim 0.03$ (snapshots \textit{a}, \textit{c}, \textit{f}), which is $\sim 8$ times larger than the intrinsic value. This failure to trigger the axisymmetric instability also takes place with a turbulent Prandtl number of $\mathrm{Pr}_\mathrm{turb}\sim 2$, while the intrinsic Prandtl number would be fixed to unity; as found by \citet{VanonOgilvie2017}, smaller Prandtl numbers further hinder the slow mode instability.  

\subsection{Disruption by non-axisymmetric instability} \label{sec:non-axi}
The stability analysis carried out in \citet{VanonOgilvie2016} however found that zonal flows of intermediate wavelengths can be disrupted by the action of a non-axisymmetric instability. Figure~\ref{fig:2zf} hints towards the presence of such an instability, as the amplitudes of the dominating wavenumbers feature moments of decay in their growth during heating events. In fact, a power spectrum map for $k_y\neq 0$ shows that the $k_y \pi G\Sigma_0/\Omega^2=0.25$ mode is also activated during heating events.

To better understand the role of this non-axisymmetric mode possessing a wavelength matching the box size, a simpler test run was conducted with two sets of ICs. In Case 1 the ICs were entirely composed of an imposed zonal flow with wavenumber $k_x \pi G\Sigma_0/\Omega^2=2.0$, making the system fully axisymmetric; in Case 2, on the other hand, the imposed zonal flow was accompanied by random velocity perturbations as described in Section~\ref{sec:ic} to provide non-axisymmetry to the system.
\begin{figure}
  \centering 
  \includegraphics[width=\columnwidth]{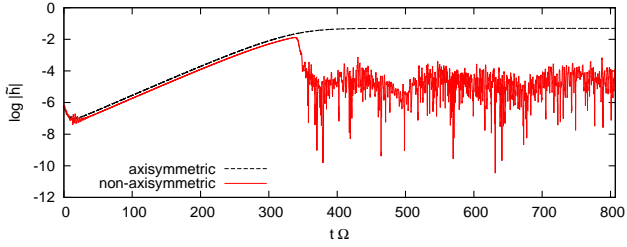}
  \caption{Temporal evolution of the natural log of the imposed zonal flow's amplitude in a test run with a purely axisymmetric nature (black, dashed line) and in one with a non-axisymmetric nature (red, full). Both test runs were conducted using $\alpha_\mathrm{init}\sim 0.004$ and $\mathrm{Pr}=3$. After the initial exponential growth period the two test runs diverge, the non-axisymmetric case showing a quenched amplitude compared to its axisymmetric counterpart.}
  \label{fig:axi-inst}
\end{figure}
The log of the amplitude of the imposed zonal flow, which grows thanks to the axisymmetric slow mode instability discussed above, is plotted in Figure~\ref{fig:axi-inst} as a function of time for both cases. While both runs show a similar exponential growth stage, they subsequently diverge: the fully axisymmetric run showing a saturation in the zonal flow amplitude (black, dashed line), and the non-axisymmetric case (red, full) exhibiting a clear amplitude quenching. This shows that some disruptive non-axisymmetric instability is present, acting to limit the maximum amplitude of the zonal flow for a given set of disc parameters (i.e. $Q$, zonal flow wavelength).

To check whether this instability is indeed the same as the one described in \citet{VanonOgilvie2016}, a $kc_s/\Omega$-$A_h$ diagram is constructed -- where $A_h$ is the zonal flow amplitude in $h$ -- using the average values from the non-axisymmetric test run (Case 2). This allows a direct comparison with the results from \citet{VanonOgilvie2016}.
\begin{figure}
  \centering
  \includegraphics[width=\columnwidth]{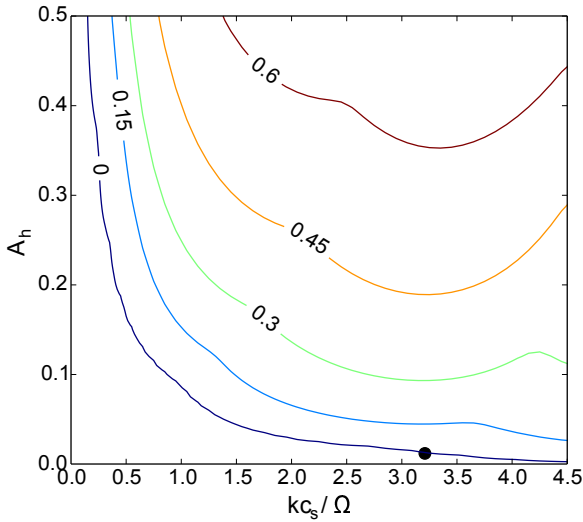}
  \caption{Growth rate contours maximised over $ky$ as a function of the zonal flow amplitude and wavenumber for $Q\simeq 1.57$ and $A_e=0$. The data point represents the average zonal flow properties obtained from the non-axisymmetric test run.}
  \label{fig:nonaxi-gen}
\end{figure}
The plot, shown in Figure~\ref{fig:nonaxi-gen}, shows the $k_y-$maximised growth rate contours as a function of the zonal flow properties (wavenumber and amplitude) for a given disc temperature (in this case $Q\simeq 1.57$ is used). The  black data point represents the test run Case 2, with the zonal flow wavenumber having been converted to acoustic units. The data point lies across the $\lambda \Omega^{-1} = 0$ contour which, taking into consideration the oscillations in $A_h$ featured in Figure~\ref{fig:axi-inst} and the fact that the wavenumber convertion to acoustic units depends on the similarly-oscillating $Q$, gives a strong indication the non-axisymmetric instability may be at play.

\begin{figure}
  \centering
  \includegraphics[width=\columnwidth]{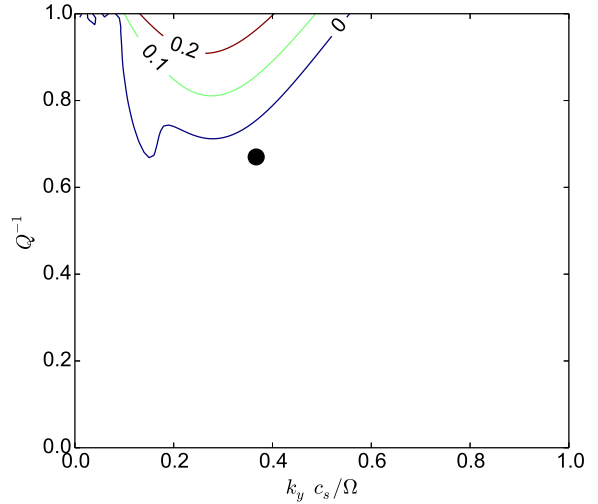}
  \caption{Growth rate contours in the $k_y c_s/\Omega$--$Q^{-1}$ domain for the average zonal flow properties from the non-axisymmetric test run: $Q\simeq 1.57$ and $A_h\simeq 0.0105$. The data point again represents the average position on the domain of the test run's zonal flow. Both shape and size of the unstable regions depend on $k$ and $A_h$.}
  \label{fig:nonaxis-ky}
\end{figure}

A more stringest test to check the activation of the non-linear instability in the gravito-turbulent regime consists in constructing a $k_y c_s/\Omega$--$Q^{-1}$ plot for the specific zonal flow wavelength ($k c_s/\Omega \simeq 3.1$, obtained from converting $k \pi G\Sigma_0/\Omega^2 = 2.0$ into acoustic units using the run's average $Q$ value) and average amplitude from the test run. The plot, shown in Figure~\ref{fig:nonaxis-ky}, illustrates to what $Q$ values the instability stretches for a given non-axisymmetric perturbation wavelength $k_y$. The data point represents $k_y \pi G\Sigma_0/\Omega^2=0.25$ converted into acoustic units, and lies close to the marginal stability contour. It is important to remember however that both the location of the data point and the shape and size of the instability region are a function of the run's parameters. The snapshot shown here is constructed using average values from the run, where the zonal flow amplitude should not be sufficiently large to trigger the non-axisymmetric instability. It is however apparent that an increase in both $Q^{-1}$ and $A_h$ would likely push the data point within the instability region. The plot also indicates that the non-axisymmetric instability would prefer a longer-wavelength mode -- which would be accessible by considering a more azimuthally elongated box -- over the $k_y \pi G\Sigma_0/\Omega^2=0.25$ mode activated with the current box parameters.

\subsection{Structure regeneration}
\begin{figure*}
  \centering
  \includegraphics[width=45mm]{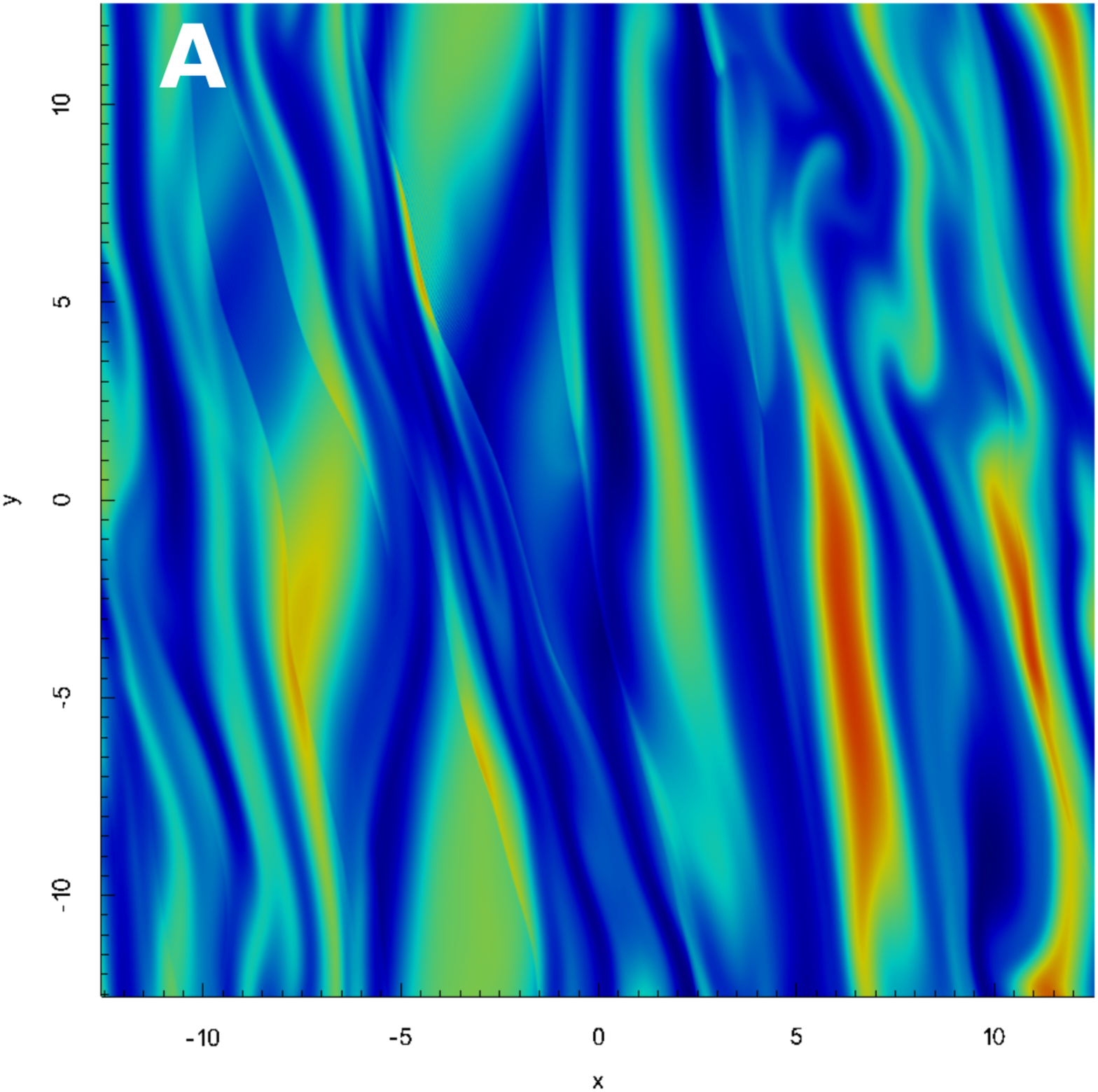}
  \includegraphics[width=45mm]{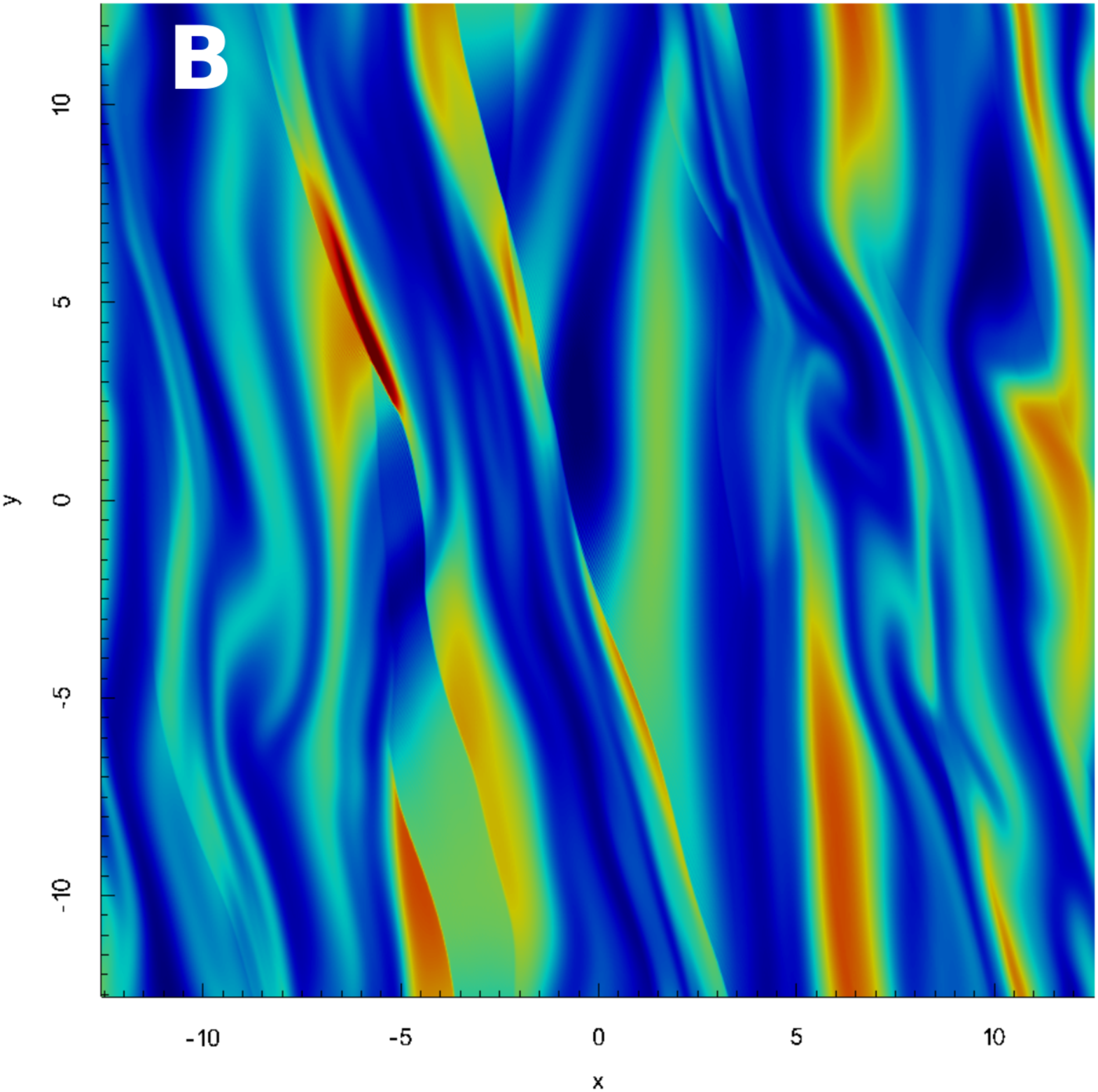}
  \includegraphics[width=45mm]{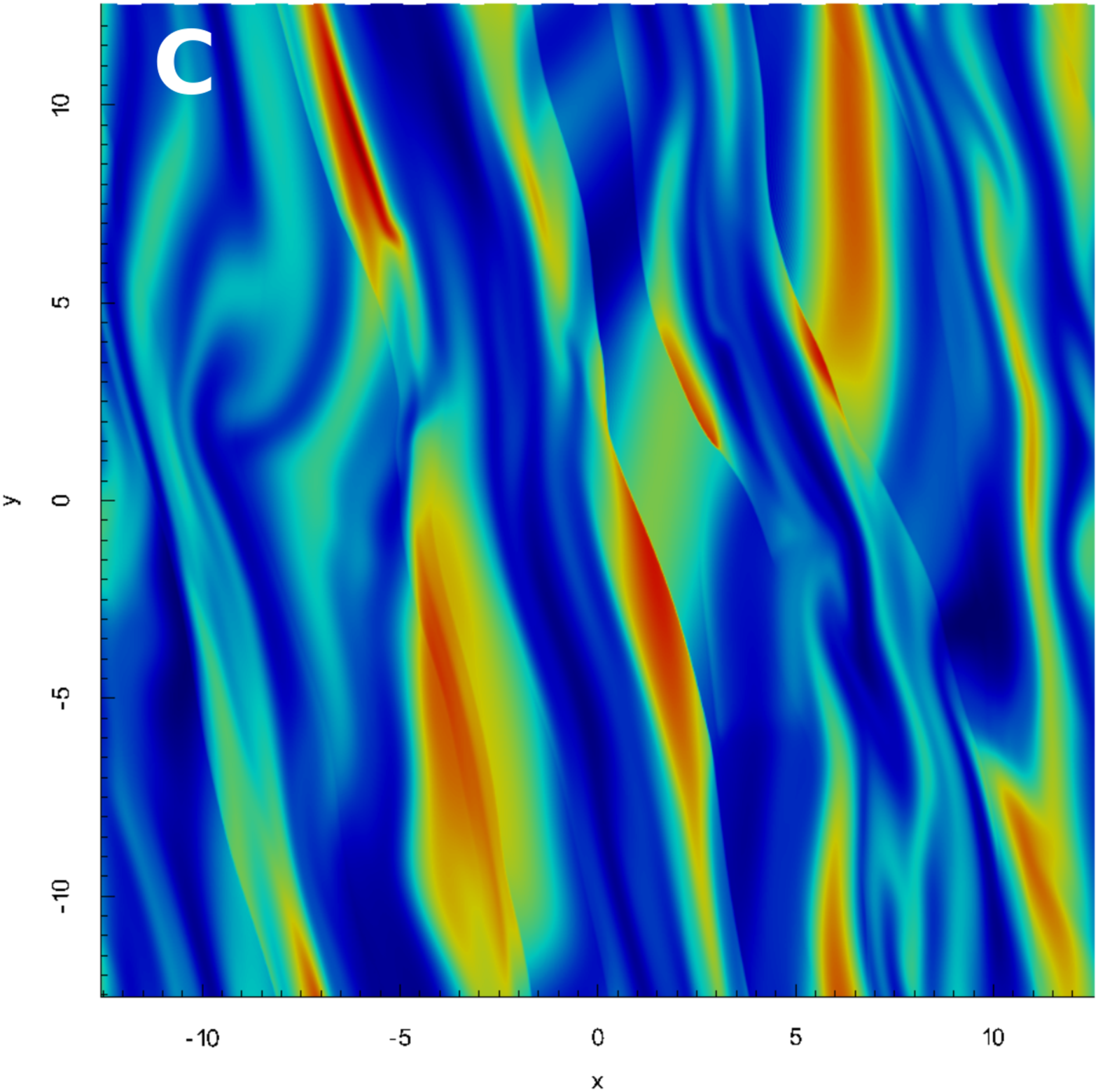}\\
  \hspace*{1cm}\includegraphics[width=45mm]{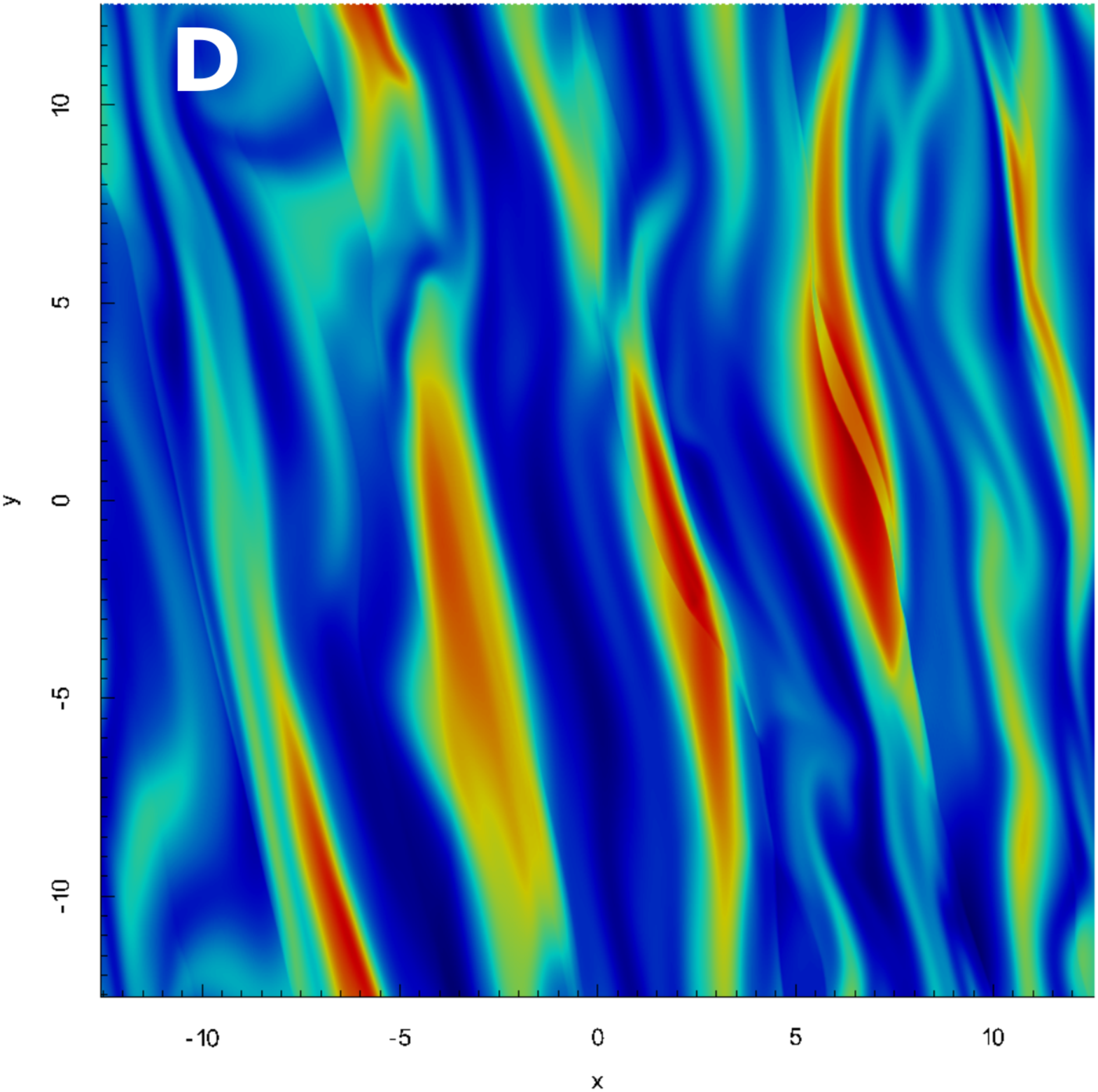}
  \includegraphics[width=45mm]{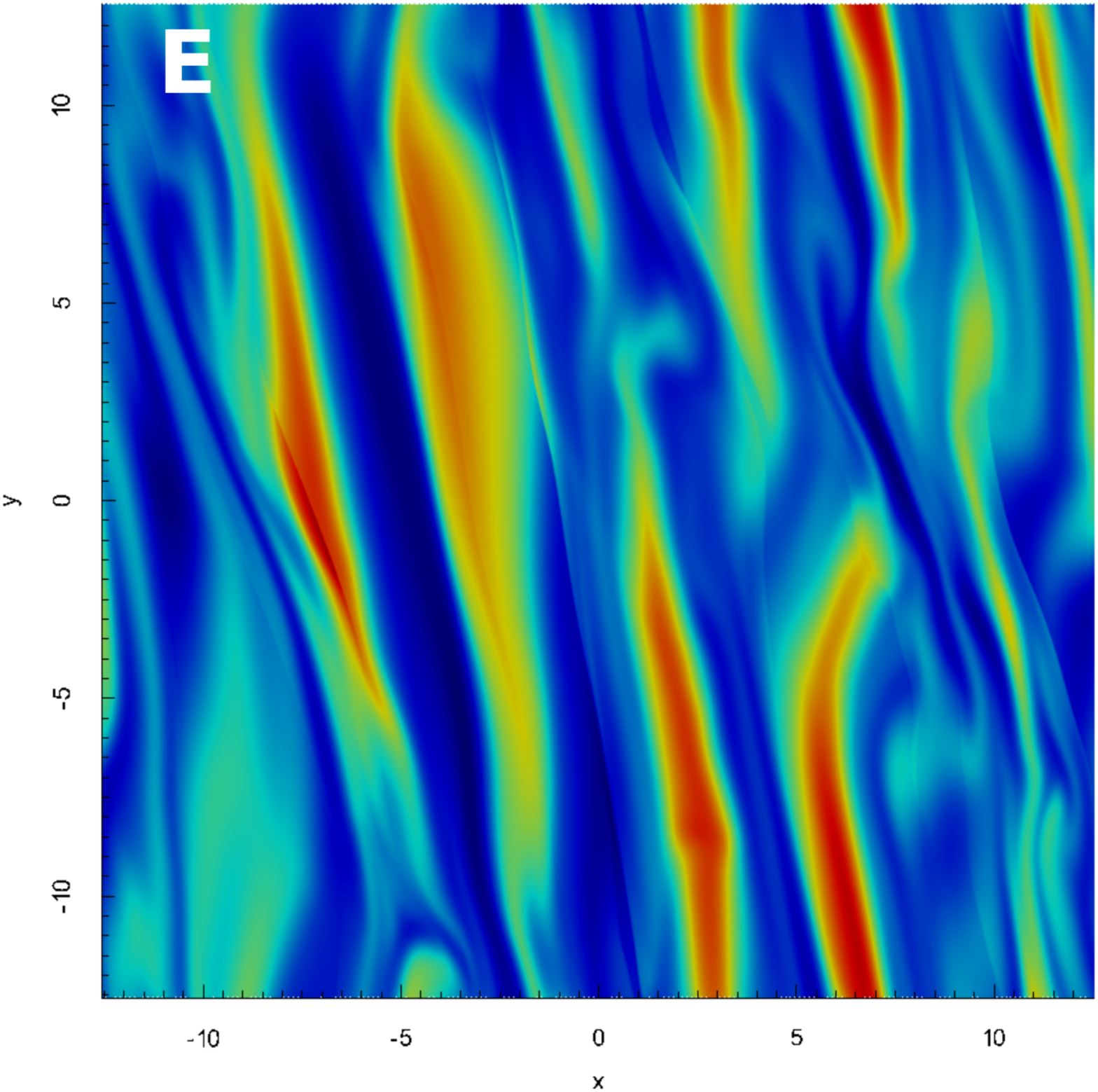}
  \includegraphics[width=55.3mm]{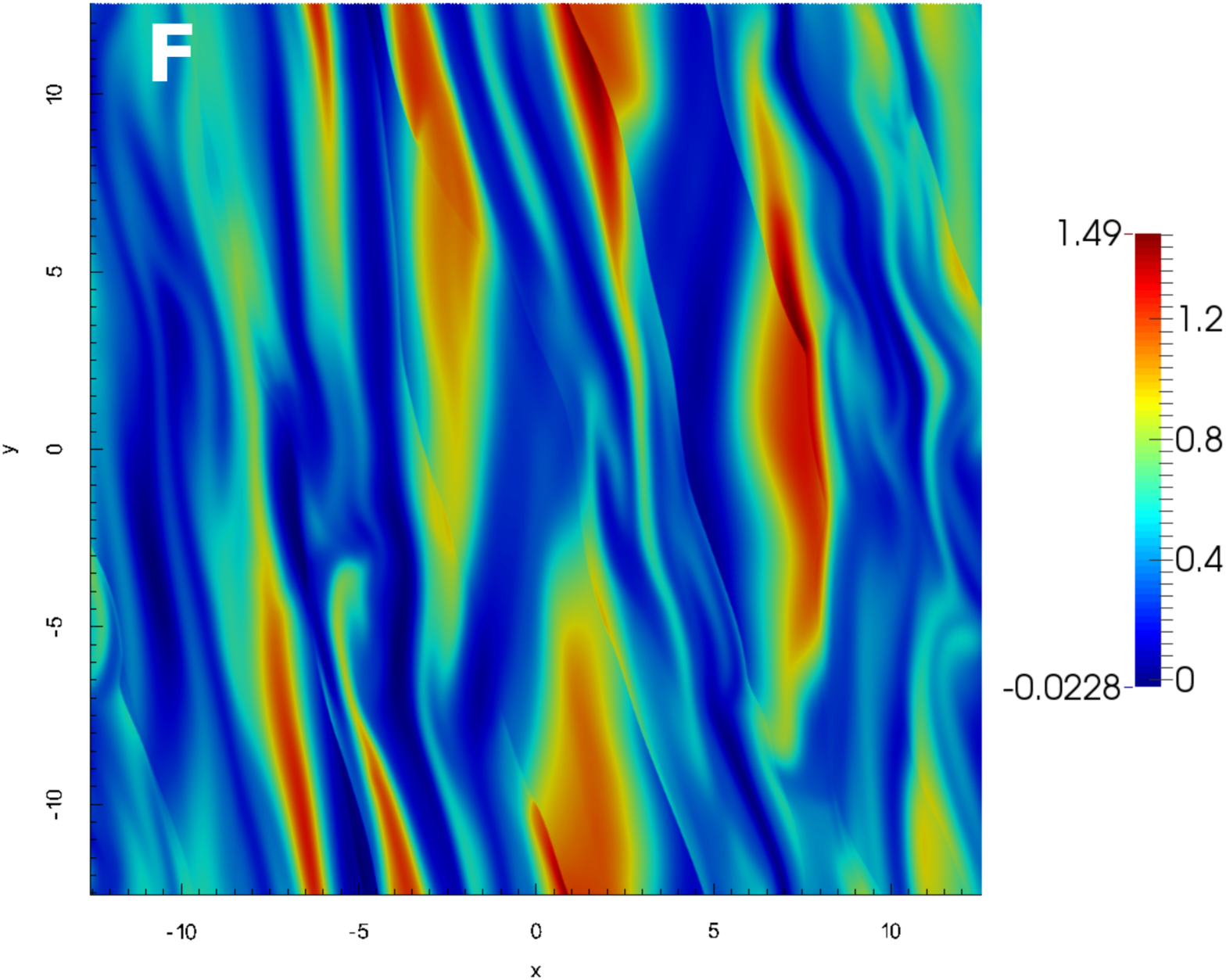}
  \caption{Entropy time sequence illustrating the destruction and regeneration of the zonal flow for $\tau_c\Omega = 10$. The total time elapsed from snapshot \textit{A} to \textit{F} is $t\Omega \simeq 4.5$.}
  \label{fig:structure-sequence}
\end{figure*}

Given the results from Sections~\ref{sec:axi} and~\ref{sec:non-axi}, it seems clear that the gravito-turbulent regime is maintained by the axisymmetric slow mode instability (whose action allows zonal flow growth) and the non-axisymmetric instability (whose action disrupts the zonal flows) balancing each other. 
This is confirmed visually by the flow dynamics in real space, a few snapshots of which are presented in Figure~\ref{fig:structure-sequence}. The snapshots, which are obtained in the entropy for a run with $\tau_c \Omega =10$, span a time of $t\Omega \simeq 4.5$.

The sequence starts (snapshot \textit{A}) with some axisymmetry on the right part of the box, where the remnants of a zonal flow linger; on the left half the zonal flow has however been disrupted, resulting in the formation of leading and trailing shearing waves. In the following two frames (\textit{B}, \textit{C}) the shearing waves steepen into shocks and merge, the merging occurring at separate times in different parts of the box due to the shocks' curved fronts. The hotspots caused by such mergers (\textit{D}) are then sheared by the flow such that a nearly axisymmetric structure forms again in most of the box (\textit{E}). The zonal flow is soon disrupted again by the non-axisymmetric instability, again resulting in the creation of leading and trailing shearing waves (\textit{F}), completing the cycle. Such a cycle repeats multiple times throughout a single heating event. Furthermore, it is possible to appreciate that the zonal flow amplitude has increased during the cycle due to the constructive effect of the slow mode instability.

The sustenance of the gravito-turbulent regime, mediated by a balance between the slow mode and non-axisymmetric instability, therefore occurs in a cyclic fashion: as the temperature in the disc drops, heat generated through viscous shock dissipation increases thanks to a growth in $\alpha_\mathrm{eff}$. This causes an enlargement in the slow mode instability region, eventually making the system unstable; as the zonal flow forms the disc cools down, as radiative cooling dominates over the scarce shock dissipation. The slow mode instability then allows the zonal flow to grow until its amplitude is large enough to trigger the destructive non-axisymmetric instability. This leads to the creation of leading and trailing shearing waves, in turn boosting the Reynolds stress which causes energy from the background flow to be fed to the turbulent motions. This boosts the kinetic energy, which is then again dissipated into heat as the shocks merge again, restarting the cycle. This cycle occurs multiple times during the course of a heating event, as well as at the beginning of each heating event.

\section{Discussion and Conclusions} \label{sec:discussion}
The work focused on the role played by zonal flows in the self-sustenance of the gravito-turbulent regime in astrophysical discs. The problem was tackled using a local shearing sheet approximation, solving the full non-linear equations of the system thanks to a bespoke pseudo-spectral method. The disc taken into consideration was assumed to be compressible, self-gravitating, viscous (with both bulk and shear viscosity types), thermally diffusive and cooled down by a constant cooling ($\beta$ cooling prescription).

The system, whose initial conditions were well out of thermal equilibrium, quickly settled into a gravito-turbulent state with an average $Q$ value of $\overline{Q}\sim 2$. The thermal balance attained by the flow in its gravito-turbulent configuration was attributed to turbulent viscosities. The onset of gravito-turbulence was accompanied by that of two axisymmetric structures, whose wavelengths remained roughly constant during the runs ($k \pi G\Sigma_0/\Omega^2 \sim 1$). Further analysis into the slow mode instability, originally discussed in \citet{VanonOgilvie2017}, showed that such an instability acted on the axisymmetric structures accompanying the gravito-turbulent regime, allowing them to grow. Such growth was however limited by the onset of a second instability, this time non-axisymmetric in nature and discussed in \citet{VanonOgilvie2016}, which disrupted the zonal flow, creating leading and trailing shearing waves in its place. 

It is this creation of shearing waves which directly led to energy being extracted from the background flow and fed back into the gravito-turbulent regime, ensuring its survival. The shearing waves were subsequently seen to steepen into shocks, merging with similar shockfronts shortly after. The hotspots created by such mergers--thanks to the shearing nature of the flow--quickly reformed zonal flows due to the resulting increase in the shear viscosity, which triggered the slow mode instability once again, completing the cycle of the zonal flow formation-destruction. Such a cycle repeats several times during the course of a single heating event, and is also at the base of the turbulent regime's ability to self-sustain on a long-term basis.

It is therefore clear that zonal flows may play an important role in the self-sustenance of a gravito-turbulent regime in the conditions described by this work, as it is ultimately their formation and destruction which allows the turbulent state to repeatedly extract energy from the background flow to maintain itself. Both slow mode and non-axisymmetric instabilities are therefore also a key part in the self-sustenance of such a state, as without the persistent zonal flow destruction by the non-axisymmetric instability, the gravito-turbulent regime would likely run out of energy.

It is however possible that other similar numerical analyses may not find comparable results if the conditions are not suitable for triggering the slow mode instability. In that case the flow would not spontaneously form axisymmetric structures, and any induced zonal flows would quickly decay. To ensure that such a cycle is set up, it is also crucial to ensure that the non-axisymmetric instability can be activated; as this instability prefers perturbations of long azimuthal wavelength, considering a box elongated in the azimuthal direction would help.

Finally, scope remains to improve the present analysis. Potential improvements could be the consideration of more refined physical approximations, improving upon simpler prescriptions such as a constant cooling timescale, or constant kinematic viscosities. Furthermore, it would be of extreme interest to test the observed self-sustaining cycle in a 3D setup.  

\section*{Acknowledgements}
The research was conducted thanks to the funding received by the Science \& Technology Facilities Council (STFC). The author would like to thank Prof. Gordon Ogilvie for the help and feedback he provided on this work, and the referee for the useful comments.

%%%%%%%%%%%%%%%%%%%%%%%%%%%%%%%%%%%%%%%%%%%%%%%%%%

%%%%%%%%%%%%%%%%%%%% REFERENCES %%%%%%%%%%%%%%%%%%

% The best way to enter references is to use BibTeX:

\bibliographystyle{mnras}
\bibliography{Vanon2017.bib} % if your bibtex file is called example.bib

% Alternatively you could enter them by hand, like this:
% This method is tedious and prone to error if you have lots of references
%\begin{thebibliography}{99}
%\bibitem[\protect\citeauthoryear{Author}{2012}]{Author2012}
%Author A.~N., 2013, Journal of Improbable Astronomy, 1, 1
%\bibitem[\protect\citeauthoryear{Others}{2013}]{Others2013}
%Others S., 2012, Journal of Interesting Stuff, 17, 198
%\end{thebibliography}

%%%%%%%%%%%%%%%%%%%%%%%%%%%%%%%%%%%%%%%%%%%%%%%%%%

%%%%%%%%%%%%%%%%% APPENDICES %%%%%%%%%%%%%%%%%%%%%

% Don't change these lines
\bsp	% typesetting comment
\label{lastpage}
\end{document}